\documentstyle[12pt,epsf]{article}
\renewcommand{\baselinestretch}{1.4}

\begin{document}
\thispagestyle{empty}
\renewcommand{\baselinestretch}{1.5}
\newcommand{\beq}{\begin{equation}}
\newcommand{\eeq}{\end{equation}}
\newcommand{\beqa}{\begin{eqnarray}}
\newcommand{\eeqa}{\end{eqnarray}}
\def\nue{{\nu_e}}
\def\anue{{\bar{\nu_e}}}
\def\numu{{\nu_{\mu}}}
\def\anumu{{\bar{\nu_{\mu}}}}
\def\nutau{{\nu_{\tau}}}
\def\anutau{{\bar{\nu_{\tau}}}}
\newcommand{\dm}{\mbox{$\Delta{m}^{2}$~}}
\newcommand{\st}{\mbox{$\sin^{2}2\theta$~}}
\def\cabsq{{c_{12}^2}}
\def\cacsq{{c_{13}^2}}
\def\cbcsq{{c_{23}^2}}
\def\sabsq{{s_{12}^2}}
\def\sacsq{{s_{13}^2}}
\def\sbcsq{{s_{23}^2}}

\begin{center}
{\large{ \bf {A three generation oscillation analysis of the Super-Kamiokande
atmospheric neutrino data beyond one mass scale dominance approximation
}}} \\
{\it  Sandhya Choubey \footnote{sandhya@tnp.saha.ernet.in},
Srubabati Goswami \footnote{sruba@tnp.saha.ernet.in},
Kamales Kar \footnote{kamales@tnp.saha.ernet.in}\\
Saha Institute of Nuclear Physics,\\1/AF, Bidhannagar,
Calcutta 700064, INDIA.\\}
\vskip 8pt


keywords: massive neutrino, mixing, atmospheric neutrinos

\vskip 6pt

{\bf{Abstract}}
\end{center}
\vskip -15pt

{\small In this paper we do a three-generation oscillation 
analysis of the latest (1144 days) Super-Kamiokande (SK)
atmospheric neutrino data going beyond the one mass scale dominance (OMSD)
approximation. We fix $\Delta_{12} = \Delta_{13}$ ($\Delta_{LSND}$) 
in the range eV$^2$ as allowed by the results from
LSND and other accelerator and reactor experiments on neutrino
oscillation and keep $\Delta_{23}$ ($\Delta_{ATM}$) and the three mixing 
angles as free parameters. 
We incorporate the matter effects, indicate some new allowed regions 
with small $\Delta_{23}$ ($< 10^{-4}$ eV$^2$) and $\sin^2 2\theta_{23}$ 
close to 0 and discuss the differences with the 
two-generation and OMSD pictures. In our scenario, 
the oscillation probabilities for the accelerator and reactor neutrinos 
involve only two of the mixing angles $\theta_{12}$ and $\theta_{13}$
and one mass scale. But the atmospheric neutrino oscillation is in general 
governed by both mass scales and all the  three mixing
angles. The higher mass scale gives rise to \dm independent average
oscillations for atmospheric neutrinos and does not enter the $\chi^2$
analysis as an independent parameter. The $\Delta_{23}$ and the three
mixing angles on the other hand appear as independent parameters in 
the $\chi^2$ analysis and the best-fit values of these are determined
from an analysis of a) the SK data, b) the SK and CHOOZ data. 
The allowed values of the mixing angles $\theta_{12}$ and $\theta_{13}$ 
from the above analysis are compared with the constraints from all
accelerator and reactor experiments including the latest results 
from LSND and KARMEN2. 
Implications for future long baseline experiments are discussed. }

\section{Introduction}

The Super-Kamiokande results on atmospheric neutrino flux measurement
show a deficit of the $\nu_\mu$ flux \cite{sk,sknew}. 
Two generation analyses of the SK
data show that the $\nu_\mu - \nu_\tau$
oscillation hypothesis provides a very good fit to the
SK data \cite{lisi,G_G,yasuda}\footnote{The $\numu-\nu_s$ 
solution is now ruled out at 99\% C.L. by the SK collaboration \cite{sknew}.}. 
The high statistics of SK also makes it possible to
study the zenith-angle dependence of the neutrino flux from which
one can conclude that the $\nu_\mu$'s show signs of oscillation but the 
$\nu_e$ events are consistent with the no-oscillation hypothesis. 
Independently the results from the reactor experiment CHOOZ  \cite{chooz} 
disfavours the 
$\nu_\mu - \nu_e$ oscillation hypothesis in a two-generation analysis.
It is important however to see the implications of these results  
in a three-generation picture. 
The most popular three-generation picture 
in the context of the SK data is the
scenario shown in fig. 1a, 
where one of the mass squared differences is in the solar neutrino
range and the other is suitable for atmospheric neutrino oscillations
\cite{lisi,yasuda3}. 
In such a scheme one mass scale dominance applies for atmospheric
neutrinos and the relevant probabilities are functions of two of the 
mixing angles and one mass squared difference. 
This picture however cannot explain the LSND results \cite{lsnd}.  
In this paper we perform a three flavor $\chi^2$-analysis of the SK 
atmospheric neutrino data assuming a mass pattern with $\Delta_{12} 
\simeq \Delta_{13}$ fixed in the eV$^2$ range and 
allowing the other mass scale to vary arbitrarily.
This mass pattern is shown in fig. 1b. 
Apart from being suitable to explain the SK atmospheric neutrino data 
this spectrum is also interesting for the laboratory based 
neutrino oscillation experiments as the higher mass scale is explorable  
in the short base line experiments, whereas the lower mass scale 
can be probed in the long base line experiments. 
In this scheme to a good approximation,
neutrino oscillation in the short-base line 
accelerators and reactors will be governed
by one (the higher) mass scale 
\cite{one,fogli94} -- and only two of the mixing
angles appear in the expressions for the oscillation probabilities.
For the atmospheric and the long baseline experiments 
the characteristic energy and length
scales are such that in general both mass differences are of relevance 
and the probabilities involve all the three mixing angles. 
However the higher mass scale gives rise to 
\dm independent average oscillations 
and it does not enter the $\chi^2$ fit directly. 
We determine the best-fit values of 
$\Delta_{23}$ and the three mixing angles by performing a $\chi^2$
analysis of 
\begin{itemize}
\item the SK atmospheric neutrino data
\item SK atmospheric and CHOOZ data
\end{itemize}
Finally we compare the allowed values of the mixing angles as obtained
from the above analysis with those allowed by the other accelerator and
reactor neutrino oscillation data including LSND and KARMEN2.

The mass scheme of this paper was first considered in \cite{minakata,ska}
after the declaration of the LSND result. 
These papers performed a combined three
generation analysis of accelerator and reactor results as well as the 
Kamiokande atmospheric neutrino data. 
Three-generation picture 
with the higher mass difference in the eV$^2$ range and the lower 
mass difference in the atmospheric range has also been considered in  
\cite{ap,flms97} (pre-SK) and \cite{thun,ol1,flms99,ol2} (post-SK). 
These papers attempted to explain both solar and 
atmospheric neutrino anomalies mainly by
maximal $\nu_{\mu} \leftrightarrow \nu_e$  oscillations driven by
$\Delta_{ATM} \sim 10^{-3}$ eV$^2$. 
Although it was claimed in \cite{thun,ol1} that this scenario can provide a
good fit to all the available data on neutrino oscillations, it was shown in
\cite{flms99} and also later in \cite{ol2} that this scenario cannot reproduce
the zenith angle dependence of the SK atmospheric neutrino data.   

{\it In this paper our aim is to
determine the allowed oscillation parameter ranges consistent 
with SK atmospheric, CHOOZ, LSND and other accelerator and 
reactor experiments.} The solar neutrino problem 
can be explained by invoking a sterile neutrino.  
We discuss in the conclusions how the solar neutrino flux
suppression can be explained 
in our scenario.

The plan of the paper is as follows. In section 2 we 
discuss very briefly 
the atmospheric neutrino code employed for the analysis of the 
SK data. In section 3 we present the 
formalism for three-generation oscillation
analysis and calculate the required probabilities including the 
earth matter effects. 
In section 4 we present the $\chi^2$ analysis 
of only SK atmospheric neutrino data. In section 5 we present the
combined $\chi^2$ analysis of SK and CHOOZ data. 
In section 6 we compare the allowed values of mixing angles from the
above analyses with those allowed by the other accelerator and
reactor data including the latest results from LSND and KARMEN2. 
In section 7 we discuss the implications of 
our results for the future long baseline experiments 
and end in section 8 with some discussions and conclusions. 

\section{The Atmospheric Neutrino Code}

We define the quantities
${N_\mu}_{osc}$ and ${N_e}_{osc}$ as
\begin{center}
${N_\mu}_{osc} = N_{\mu\mu} + N_{e\mu}$\\
${N_e}_{osc} = N_{ee} + N_{\mu e}$
\end{center}
${N_{e,\mu}}_{osc}$ are the numbers of $e$-like and
$\mu$-like events in the detector and 
$N_{l l^\prime}$ is defined as 
\begin{eqnarray}
N_{l l^{\prime}}
& = & n_T
\int^{\infty}_{0} dE
\int^{(E_{l^{\prime}})_{\rm max}}_{(E_{l^{\prime}})_{min}} dE_{l^{\prime}}
\int_{-1}^{+1} d\cos \psi
\int_{-1}^{+1} d\cos \xi\
{1 \over 2\pi}
\int_{0}^{2\pi} d\phi
\nonumber\\
&\times&
{d^2F_l (E,\xi) \over dE~d\cos\xi}
\cdot{ d^2\sigma_{l^{\prime}} (E,E_{l^{\prime}},\cos\psi) \over
dE_{l^{\prime}}~d\cos\psi }  
\epsilon(E_{l^{\prime}})
\cdot
{\ }P_{\nu_l \nu_{l^{\prime}}} (E, \xi).
\label{rate}
\end{eqnarray}
$n_T$ denotes the number of target nucleons, $E$ is the neutrino energy, 
$E_{l^{\prime}}$ is the energy of the final charged lepton, $\psi$ is
the angle  
between the incoming neutrino $\nu_l$ and the scattered lepton $l^{\prime}$, 
$\xi$ is the zenith angle of the neutrino and $\phi$ is the azimuthal angle 
corresponding to the incident neutrino direction.  
The zenith angle of the
charged lepton is given by
\begin{equation}
\cos \Theta = \cos \xi \cos \psi + \sin \xi \cos \phi \sin \psi
\label{zenith}
\end{equation}
$d^2\sigma_{l^{\prime}}/dE_l d\cos\psi$ is the differential cross section
for $\nu_{l^{\prime}} N \rightarrow l^{\prime}  X$ scattering, 
$\epsilon(E_{l^\prime})$  
is the detection efficiency for the 1 ring events in the detector and 
$P_{\nu_l \nu_{l^\prime}}$ is the  probability of a 
neutrino flavour $l$ to convert to a neutrino of flavour 
$l^{\prime}$. 
We use the atmospheric neutrino fluxes 
$d^2F_l (E,\xi) \over dE~d\cos\xi$ from \cite{honda}. 
For further details regarding the calculation of number of events
we refer to \cite{ssdecay}.

\section{Three-Flavor Analysis}

\subsection{The vacuum oscillation probabilities}
The general expression for the probability that an initial
$\nu_{\alpha}$ of energy $E$ gets converted to a $\nu_{\beta}$
after traveling  a distance $L$ in vacuum is given by, 
\begin{equation}
P(\nu_{\alpha},0 ; \nu_{\beta},t) =  \delta_{\alpha \beta} - 4~
\sum_{j > i}~ U_{\alpha i} U_{\beta i} U_{\alpha j} U_{\beta j}
\sin^{2}\left(\frac{\pi L}{\lambda_{ij}}\right)
\label{npr}
\end{equation}
where 
$\lambda_{ij}$ is defined to be the neutrino vacuum oscillation
wavelength given by,  
\begin{equation}
\lambda_{ij} = (2.47 {\rm m}) \left(\frac{E}{MeV}\right)
\left(\frac{\rm {eV}^2}{\Delta_{ij}}\right) 
\label{wv}
\end{equation}
which denotes the scale over which neutrino oscillation effects can be 
significant and  
$\Delta_{ij} = \mid{m_{j}^2 - m_{i}^2}\mid$.
The actual forms of the various survival and transition
probabilities depend on the spectrum of ${\Delta m}^{2}$ assumed
and the choice of the mixing matrix $U$ relating the flavor
eigenstates to the mass eigenstates. We choose the flavor states
$\alpha =$ 1,2, and 3 to correspond to e, $\mu $ and $\tau $
respectively.   The most suitable
parameterization of $U$ for the mass spectrum chosen by us is $U =
R_{13} R_{12} R_{23}$ where $R_{ij}$ denotes the rotation matrix in
the $ij$-plane. This yields:
\begin{equation}
U = {\pmatrix {c_{12}c_{13} & s_{12}c_{13}c_{23} - s_{13}s_{23}
& c_{13}s_{12}s_{23} + s_{13}c_{23} \cr
-s_{12} & c_{12}c_{23} &c_{12}s_{23} \cr
-s_{13}c_{12} & -s_{13}s_{12}c_{23} - c_{13}s_{23} &
-s_{12}s_{13}s_{23} + c_{13}c_{23}\cr}}
\label{um}
\end{equation}
where $c_{ij} =\cos{{\theta}_{ij}}$ and $s_{ij}
=\sin{{\theta}_{ij}}$ here and everywhere else in the paper. 
We have assumed CP-invariance so that $U$ is
real. The above choice of $U$ has the advantage that ${\theta}_{23}$
does not appear in the expressions for the probabilities for the
laboratory experiments \cite{ska}. 

\noindent
The probabilities relevant for atmospheric neutrinos are
\renewcommand{\theequation}{6\alph{equation}}
\setcounter{equation}{0}
\begin{eqnarray}
P_{\nu_e \nu_e}&=&1 - 2c_{13}^{2}c_{12}^2 + 2c_{13}^{4}c_{12}^4 - 4
(c_{13}s_{12}c_{23} - s_{13}s_{23})^{2} (c_{13}s_{12}s_{23} +
s_{13}c_{23})^{2}
~\rm S_{23}\\
\label{pnueatm} 
P_{\nu_\mu \nu_e}&=&2c_{13}^2c_{12}^2s_{12}^2 - 4c_{12}^2c_{23}s_{23}
(c_{13}s_{12}c_{23} - s_{13}s_{23})(c_{13}s_{12}s_{23} +
s_{13}c_{23}) 
~\rm S_{23}\\
\label{pmueatm}
P_{\nu_\mu \nu_\mu}&=&1 - 2c_{12}^{2}s_{12}^2 -
4c_{12}^4c_{23}^2s_{23}^2
~\rm S_{23}
\label{pnumuatm}
\end{eqnarray} 
\setcounter{equation}{6}
\renewcommand{\theequation}{\arabic{equation}}
\noindent 
where $\rm S_{23} = \sin^2(\pi L/\lambda_{23})$. 
Apart from the most general three generation regime, the following 
limits are of interest, as we will see later in the context of the SK 
data:
\begin{enumerate}
\item  {The two-generation limits}\\
Because of the presence of more parameters as compared to the one mass
scale dominance picture there are twelve possible two-generation 
limits \cite{sg} with the oscillations driven by either $\Delta_{LSND}$
or $\Delta_{ATM}$. Below we list these limits specifying the mass scales that 
drive the oscillations:
\begin{itemize}
\item{
$s_{12} \rightarrow 0, s_{13} \rightarrow 0~~~~~ (\nu_\mu-\nu_\tau,~~ \Delta_{ATM}$) \\
$s_{12} \rightarrow 1, s_{13} \rightarrow 0~~~~~ (\nu_e -\nu_\tau,~~ \Delta_{ATM}$)\\
$s_{12} \rightarrow 0, s_{13} \rightarrow 1~~~~~ (\nu_\mu-\nu_e,~~ \Delta_{ATM}$)\\
$s_{12} \rightarrow 1, s_{13} \rightarrow 1~~~~~ (\nu_e-\nu_\tau,~~ \Delta_{ATM}$)}

\item{
$s_{13} \rightarrow 0, s_{23} \rightarrow 0~~~~~(\nu_\mu - \nu_e,~~ 
\Delta_{LSND}$)\\
$s_{13} \rightarrow 0, s_{23} \rightarrow 1~~~~~(\nu_\mu - \nu_e,~~ 
\Delta_{LSND}$)\\  
$s_{13} \rightarrow 1, s_{23} \rightarrow 0~~~~~(\nu_\mu - \nu_\tau,~~ 
\Delta_{LSND}$)\\
$s_{13} \rightarrow 1, s_{23} \rightarrow 1~~~~~(\nu_\mu - \nu_\tau,~~ 
\Delta_{LSND}$)
 }

\item{
$s_{12} \rightarrow 0, s_{23} \rightarrow 0~~~~~(\nu_e-\nu_\tau,
~~ \Delta_{LSND}$)\\
$s_{12} \rightarrow 0, s_{23} \rightarrow 1~~~~~(\nu_e - \nu_\tau,~~ 
\Delta_{LSND}$)\\
$s_{12} \rightarrow 1, s_{23} \rightarrow 0~~~~~(\nu_e-\nu_\tau,~~ 
\Delta_{ATM}$)\\
$s_{12} \rightarrow 1, s_{23} \rightarrow 1~~~~~ (\nu_e - \nu_\tau, ~~
\Delta_{ATM}$)}

\end{itemize} 

\item  
$s_{12}^2$ = 0.0 \\
In this limit the relevant probabilities become
\renewcommand{\theequation}{7\alph{equation}}
\setcounter{equation}{0}
\begin{eqnarray}
P_{\nue \nue} &=& 1 - 2 c_{13}^2 s_{13}^2 +
4 s_{13}^2 c_{23}^2 s_{23}^2 S_{23} \\
P_{\nue \numu} &=& 4 s_{13}^2 s_{23}^2 c_{23}^2 S_{23}\\
P_{\numu \numu} &=& 1 - 4 c_{23}^2 s_{23}^2 S_{23}
\end{eqnarray}
\setcounter{equation}{7}
\renewcommand{\theequation}{\arabic{equation}}
\noindent 
Thus $P_{\numu \numu}$ is the same as the two generation limit, 
$P_{\numu \nue}$ is governed by two of the mixing angles and
one mass scale and $P_{\nue \nue}$ is governed by two mixing angles and both
mass scales. 

\item 
$s_{13}^2$ = 0.0 \\
For this case the probabilities take the form 
\renewcommand{\theequation}{8\alph{equation}}
\setcounter{equation}{0}
\begin{eqnarray}
P_{\nue \nue} &=& 1 - 2 c_{12}^2 s_{12}^2 -
4 s_{12}^4 c_{23}^2 s_{23}^2 S_{23}\\
P_{\nue \numu} &=& 2 c_{12}^2 s_{12}^2 - 4 c_{12}^2 s_{12}^2 c_{23}^2 s_{23}^2
S_{23}\\
P_{\numu \numu} &=& 1 - 2c_{12}^2 s_{12}^2 - 
4 c_{12}^4 c_{23}^2 s_{23}^2 S_{23}
\end{eqnarray}
\setcounter{equation}{8}
\renewcommand{\theequation}{\arabic{equation}}
\noindent
In this case the probabilities are governed by two mass scales and two mixing angles. 

\end{enumerate}
We note that for cases (2) and (3) the probabilities are symmetric
under the transformation $\theta_{23} \rightarrow \pi/2 - \theta_{23}$.
The probabilities for these cases are functions of at most two mixing angles 
as in the OMSD case \cite{lisi} but they are governed by both mass scales
making these limits different from the OMSD limit.  

\subsection {Earth matter effects}

Since on their way to the detector the upward going neutrinos pass through the 
earth, it is important in general to include the matter effect in 
the atmospheric neutrino analysis. 
The matter contribution to the effective squared mass of the electron 
neutrinos:
\begin{eqnarray}
A = 2{\sqrt 2}~ G_F~ E~n_e 
\end{eqnarray}
where $E$ is the neutrino energy and $n_e$ is the ambient electron 
density. Assuming a typical
density of 5 gm/cc and $E$ = 10 GeV, the matter potential $A 
\simeq  3.65 \times 10^{-3}$ eV$^2$ and since this is of the 
same order as $\Delta_{23}$ in our case, 
matter effects should be studied carefully.

\noindent
The mass matrix in the flavor basis in presence of matter is given by
\begin{eqnarray}
M_F ^2 = U~M^2~U^\dagger ~+~ M_A
\end{eqnarray}
where $M^2$ is the mass matrix in the mass eigenbasis, $U$ is the 
mixing matrix and 
\begin{eqnarray}
M_A = {\pmatrix {A & 0 & 0 \cr
0 & 0 & 0 \cr
0 & 0 & 0 \cr}}
\end{eqnarray}
Since $\Delta_{12} \sim \Delta_{13} \gg \Delta_{23} \sim A$, one can 
solve the eigenvalue problem
using the degenerate perturbation theory, where the $\Delta_{23}$ 
and $A$ terms are treated as a perturbation to the dominant 
$\Delta_{12}$ and $\Delta_{13}$ dependent terms. The mixing angle in matter 
is then given by
\begin{eqnarray}
\tan 2\theta_{23}^M = \frac{\Delta_{23}\sin2\theta_{23} -
As_{12}\sin2\theta_{13}}{\Delta_{23} \cos 2\theta_{23} -
A(s_{13}^2 - c_{13}^2 s_{12}^2)}
\label{theta23m}
\end{eqnarray}
while the mass squared difference in matter turns out to be
\begin{eqnarray}
\Delta_{23}^M = {\left[(\Delta_{23} \cos 2\theta_{23} - 
A(s_{13}^2 - c_{13}^2 s_{12}^2))^2 + (\Delta_{23}\sin2\theta_{23}
- As_{12}\sin2\theta_{13})^2\right]}^{1/2}
\label{del23m}
\end{eqnarray}
The mixing angles $\theta_{12}$ and $\theta_{13}$ as well as the 
larger mass squared difference $\Delta_{12}$ remain unaltered 
in matter. From eq. (\ref{theta23m}) and (\ref{del23m}) 
we note the following 
\begin{itemize}
\item In the limit of both $s_{12}\rightarrow 0$ 
and $s_{13}$ $\rightarrow 0$, 
the matter effect vanishes and we recover the two-generation 
$\nu_{\mu} - \nu_{\tau}$ limit. 
\item The resonance condition now becomes 
$\Delta_{23}\cos 2\theta_{23} = A (s_{13}^2 - c_{13}^2 s_{12}^2)$.
So that for $\Delta_{23} > 0$, one can have resonance for both 
neutrinos -- if 
$s_{13}^2 > c_{13}^2 s_{12}^2$ -- as well as for antineutrinos -- if 
$s_{13}^2 < c_{13}^2 s_{12}^2$.
This is different from the OMSD picture where for 
$\Delta m^2 > 0$ only neutrinos can resonate \cite{uma}.

\item In the limit of $s_{12} \rightarrow 0 $ 
\begin{equation}
\tan 2\theta_{23}^{M} = \frac{\Delta_{23} \sin2\theta_{23}}{\Delta_{23} 
\cos 2\theta_{23} - A s_{13}^2}
\label{l1}
\end{equation} 
Here one gets resonance for neutrinos only (if $\Delta_{23}>0$) and 
this is similar to the OMSD case.

\item In the limit $s_{13} \rightarrow 0$ 
\begin{equation} 
\tan 2\theta_{23}^{M} = \frac{\Delta_{23} \sin 2\theta_{23}}{\Delta_{23} 
\cos 2\theta_{23} + A s_{12}^2}
\label{l2}
\end{equation} 
For this case for $\Delta_{23}>0$, there is no resonance for 
neutrinos but antineutrinos can resonate. 

\item In the limit where $\Delta_{23}\rightarrow 0$ 
\begin{eqnarray}
\tan 2\theta_{23}^M = \frac{s_{12} \sin 2\theta_{13}}
{s_{13}^2 - c_{13}^2 s_{12}^2},~~~ 
\Delta_{23}^M = A(s_{13}^2 + c_{13}^2 s_{12}^2)
\label{lowdelmsq}
\end{eqnarray}
Thus even for small values of $\Delta_{23} < 10^{-4}$ 
the mass squared difference in matter is $\sim A$ and one may 
still hope to see oscillations for the upward neutrinos due to matter
effects. The other point to note is that the mixing angle in matter 
$\theta_{23}^M$ depends only on $\theta_{12}$ and
$\theta_{13}$ and is independent of the vacuum mixing angle 
$\theta_{23}$ and $\Delta_{23}$. 
Contrast this with the OMSD case (where the expressions 
for $\tan 2\theta_{23}^M$ is given by an expression similar to 
eq. (\ref{l1}) \cite{uma}) and the two-generation 
$\nu_\mu - \nu_e$ oscillations.
For both the two-generation $\nu_\mu - \nu_e$
as well as the three-generation OMSD case, 
for $\Delta_{23} \rightarrow 0$, the mixing angle 
$\tan 2\theta_{23}^M \rightarrow 0$, 
but for the mass spectrum considered in this paper the 
$\tan 2\theta_{23}^M$ maybe large depending on the values 
of $\sabsq$ and $\sacsq$.
Hence we see that the {\it demixing effect} which
gives the lower bound on allowed values of \dm in
the two generation $\nu_\mu - \nu_e$ or the three-generation
OMSD case, does not arise here and we hope to get allowed regions 
even for very low values of $\Delta_{23}$. 
On the other hand even small values
of $\theta_{23}$ in vacuum can get enhanced in matter.
This special case where $\Delta_{23} \sim 0$ was 
considered in an earlier paper \cite{rujula}.


\item In the limit of $\sbcsq \rightarrow 0$ 
\begin{eqnarray}
\tan 2 \theta_{23}^M = \frac{-A s_{12} \sin 2\theta_{13}}
{\Delta_{23} -A (s_{13}^2 - c_{13}^2 s_{12}^2)}
\label{lowtheta1}
\end{eqnarray}
\item While for $\sbcsq \rightarrow 1$
\begin{eqnarray}
\tan 2 \theta_{23}^M = \frac{-A s_{12} \sin 2\theta_{13}}
{-\Delta_{23} -A (s_{13}^2 - c_{13}^2 s_{12}^2)}
\label{lowtheta2}
\end{eqnarray}
For the last two cases, corresponding to $\sin^2 2\theta_{23} \rightarrow 0$, 
again the mixing angle $\theta_{23}$ in matter is
independent of its corresponding value in vacuum and hence for
appropriate choices of the other three parameters, $\Delta_{23}$,
$\sabsq$ and $\sacsq$, one can get large values for $\sin^2 2\theta_{23}^M$
even though the vacuum mixing angle is zero.
  
\end{itemize}

\noindent
The amplitude that an initial $\nu_\alpha$ of energy $E$ is 
detected as $\nu_\beta$ after traveling through the earth is
\begin{eqnarray}
A(\nu_\alpha,t_0,\nu_\beta,t) = \sum_{\sigma,\lambda,\rho}
\sum_{i,j,k,l}&& [(U_{\beta l}^{M_m} e^{-iE_l^{M_m}(t-t_3)}
U_{\sigma l}^{M_m})(U_{\sigma k}^{M_m} e^{-iE_k^{M_c}(t_3-t_2)}
U_{\lambda k}^{M_c}) \times
\nonumber \\
&& (U_{\lambda j}^{M_m} e^{-iE_j^{M_m}(t_2-t_1)}U_{\rho j}^{M_m})
(U_{\rho i}e^{-iE_i (t_1-t_0)}U_{\alpha i})]
\label{amp}
\end{eqnarray}
where we have considered the earth to be made of two slabs, a mantle 
and a core with constant densities of 4.5 gm/cc and 11.5 gm/cc 
respectively and include the non-adiabatic effects at the boundaries. 
The mixing matrix in the mantle and the core are given 
by $U^{M_m}$ and $U^{M_c}$ respectively. $E_i^{X} 
\approx m_{iX}^2/2E$, $X$ = core(mantle) and
$m_{iX}$ is the mass of the 
$i^{th}$ neutrino state in the core(mantle).
The neutrino is produced at time $t_0$, hits the earth mantle at 
$t_1$, hits the core at $t_2$, leaves the core at $t_3$ and 
finally hits the detector at time $t$. The Greek indices 
($\sigma,\lambda,\rho$) denote the flavor eigenstates 
while the Latin indices ($i,j,k,l$) give the 
mass eigenstates.
The corresponding 
expression for the probability is given by
\begin{eqnarray}
P(\nu_\alpha,t_0,\nu_\beta,t) = |A(\nu_\alpha,t_0,\nu_\beta,t)|^2
\label{pr}
\end{eqnarray}
For our calculations of the number of events we have used the full 
expression given by eq.(\ref{amp}) and (\ref{pr}).

\section{$\chi^2$-analysis of the SK data}

We minimize the $\chi^2$ function defined as \cite{lisi,G_G}
\begin{equation}
\chi^2 =
\sum_{i,j=1,40} \left(N_i^{th} -
N_i^{exp}\right)
(\sigma_{ij}^{-2}) \left(N_j^{th} - N_j^{exp}\right)
\label{chi}
\end{equation}
where the sum is over the sub-GeV and multi-GeV electron and muon bins. 
The experimentally observed number of events 
are denoted by the superscript ``exp"
and the theoretical predictions for the quantities are labeled by ``th".
The element of the error matrix $\sigma_{ij}$ is calculated as in 
\cite{lisi}, including the correlations between the different bins. 
For contained events there are  forty experimental data points.
The probabilities for
the atmospheric neutrinos are explicit 
functions of one mass-squared difference and
three mixing angles making the number of degrees of freedom (d.o.f) 36. 
The other mass squared difference gives rise to \dm independent 
average oscillations and hence does not enter the fit as an 
independent parameter. 

\noindent
For two-flavour $\nu_\mu - \nu_\tau$ oscillation the 1144 days of
data gives the following best-fits and $\chi^2_{min}$:
\begin{itemize}
\item{
$\chi^2_{min}/d.o.f. = 36.23/38$, 
$\Delta m^2$ = 0.0027 eV$^2$, $\sin^2 2\theta$ = 1.0}
\end{itemize}
This corresponds to a goodness of fit of 55.14\%.

\noindent
For the general three-generation scheme
the $\chi^2_{min}$ and the best-fit values of parameters that we get are
\begin{itemize}
\item
$\chi^2_{min}/d.o.f. = 34.65/36$, $\Delta_{23} = 0.0027$ eV$^2$,
$s^2_{23} = 0.51$,$s_{12}^2 = 0.04$ and
$s_{13}^2 =0.06 $
\end{itemize}
This solution is allowed at 53.28\% C.L.

The  solid(dashed) lines in fig. 2 present 
the variation of the $\Delta \chi^2 = \chi^2 - \chi^2_{min}$ 
for the SK data, with respect to 
one of the parameters keeping the other three unconstrained, 
when we include(exclude) the matter effect.  
In fig. 2(a) as we go towards smaller values of 
$\Delta_{23}$  around $10^{-3}$ eV$^2$
the effect of matter starts becoming important   
as the matter 
term is now comparable to the mass term. 
If matter effects are not there then for values of $\Delta_{23}$ 
$\stackrel{<}{\sim} 10^{-4} $ eV$^2$ the 
$S_{23}$ term in eq.(6)
is very small and there is no up-down asymmetry
resulting in very high values of $\chi^2$ as is evident 
from the dashed curve. 
If the matter effects are included, then in the limit of very 
low $\Delta_{23}$ the matter term dominates and 
$\Delta_{23}^M$ is given by eq.(\ref{lowdelmsq}).
Since this term $\sim 10^{-3}$ eV$^2$ there can be 
depletion of the 
neutrinos passing through the earth causing an updown asymmetry. 
For $\Delta_{23}$ around $10^{-4}$ eV$^2$, there is cancellation 
between the two comparable terms in the numerator of 
eq. (\ref{theta23m}) and 
the mixing angle becomes very small and hence the $\chi^2$ around 
these values of $\Delta_{23}$ comes out to be very high. 

Fig 2(b) illustrates the corresponding variation of $\Delta \chi^2$ with
$\sbcsq$ while the other three parameters are allowed to vary 
arbitrarily. 
For small and large values of $s_{23}^2$  
the inclusion of matter effect makes a difference. 
For $s_{23}^2$ either very small or large ($\sin^2 2\theta_{23} 
\rightarrow 0$) the overall suppression of the 
$\nu_\mu$ flux is less than that required by the data if vacuum oscillation 
is operative and so it is ruled out. If we include matter effects then    
in the limit of $s_{23}^2 =0$ and $s_{23}^2 =1$ the matter mixing angle is 
given by eqs. (\ref{lowtheta1}) and (\ref{lowtheta2}), which can 
be large for suitable values of $\sacsq$ and $\sabsq$ and hence 
one gets lower $\chi^2$ even for these values of $s_{23}^2$. 

In figs 2(c) and 2(d) we show the effect of $\sabsq$ and $\sacsq$
respectively on $\Delta \chi^2$. From the solid and the dashed lines 
it is clear that matter effects do not vary much the allowed ranges of 
$\sabsq$ and $\sacsq$.    

The dashed-dotted line in the figure 
shows the 99\% C.L. (= 13.28 for 4 parameters) 
limit. In Table 1 we give the allowed ranges of the mixing 
parameters, inferred from fig. 2 
at 99\% C.L. for the SK atmospheric data, with 
and without matter effects. 
\begin{description}
\item{\bf Table 1:} The allowed ranges of parameters for the SK data.
\end{description}
\[
\begin{tabular}{|c|c|c|c|c|}\hline
{} & {$\Delta_{23}$ in eV$^2$}&{$\sbcsq$}&{$\sabsq$}&{$\sacsq$}\\ \hline
{with} & {$\!\!1.6\!\!\times \!10^{-4}\!\leq\! 
\Delta_{23}\! \leq \!7.0\!\!\times\!\! 10^{-3}\!\!$} & 
{$0.26 \leq \sbcsq \leq 0.77$} & {$\sabsq \leq 0.21$} & 
{$\sacsq \leq 0.55$} \\
{$\!\!$matter effects$\!\!$} & {$\Delta_{23} \leq 6.5\times 10^{-5}$}
& {$\sbcsq \geq 0.85$} & {} & {} \\ \hline
{without}&{$\!\!5\!\!\times\!\! 10^{-4}\!\leq \!\Delta_{23}\! 
\leq 7.0\! \times \!\!10^{-3}\!\!$}
&{$0.27 \leq \sbcsq \leq 0.74$}&{$\sabsq \leq 0.21$}&
{$\sacsq \leq 0.6$}\\
{$\!\!$matter effects$\!\!$}&{}&{}&{}&{}\\ \hline
\end{tabular}
\]

\subsection{Zenith-Angle distribution}

Since the probabilities in our case are in general 
governed by two mass scales and
all  three
mixing  angles it is difficult to understand the allowed regions.  
To facilitate the qualitative understanding we present 
in fig. 3 the histograms which describe the zenith angle distribution.  
The event distributions in these histograms are approximately given by, 
\begin{equation}
\frac{N_\mu}{N_{\mu_{0}}} \approx P_{\numu \numu} +
\frac{N_{e_{0}}}{N_{\mu_{0}}} 
P_{\nue \numu}
\label{muno}
\end{equation}
\begin{equation}
\frac{N_e}{N_{e_{0}}} \approx P_{\nue \nue} + \frac{N_{\mu_{0}}}{N_{e_{0}}} 
P_{\numu \nue}
\label{eno}
\end{equation}
where the quantities with suffix 0 indicates the no-oscillation values. 
For the sub-GeV data $N_{\mu_{0}}/N_{e_{0}} \approx 2$ to a good approximation 
however for the multi-GeV data this varies in the range 2 (for  
$\cos\Theta$ =0) to 3 (for $\cos\Theta = \pm $1) \cite{lisi}.
 

In fig. 3a we study the effect of varying $s_{12}^2$ and $s_{13}^2$
for fixed values of $\Delta_{23}$ = 0.002 eV$^2$ 
and $s_{23}^2$ = 0.5.  
From eq. (\ref{theta23m}), (\ref{del23m}) and from fig. 2 
we see that for the values of the $\Delta_{23}$ and $s_{23}^2$ 
considered in this figure 
the matter effects are small and we can understand the 
histograms from the vacuum oscillation probabilities.  
The thick solid line shows the event distribution for $s_{12}^2 = 0$
and $s_{13}^2 = 0.1$. 
As $s_{13}^2$ increases from 0, keeping $s_{12}^2$ as 0, from eqs. (9) 
$P_{\nue \nue}$ decreases from 1 and 
$P_{\nue \numu}$ increases from zero  resulting in a net
electron  depletion according to eq. (\ref{eno}). 
The long dashed line corresponds to $s_{13}^2$ = 0.3  
for which the electron depletion is 
too high as compared to data. 
The muon events are also affected as $P_{\numu \nue}$ increases with 
increasing $s_{13}^2$ even though $P_{\numu \numu}$ 
is independent of $s_{13}^2$. 
On the other hand for $s_{13}^2$ = 0.0, the effect of  
increasing $s_{12}^2$ is to increase the number of electron events 
and decrease the number of muon events according to eqs. (10), (\ref{eno}) 
and (\ref{muno}). 
This is shown by the short-dashed and dotted lines in fig. 3a. 
For $s_{12}^2$ = 0.2 the electron excess and muon 
depletion both becomes too high as compared to the data.  
For the case when both
$s_{12}^2$ and $s_{13}^2$ are 0.1 the electron depletion caused by
increasing $s_{12}^2$ and the excess caused by increasing $s_{13}^2$ 
gets balanced and the 
event distributions are reproduced quite well, 
shown by the dashed-dotted line. 

 
In fig. 3b we study the effect of varying $s_{23}^2$ and $\Delta_{23}$ 
in the limit of $s_{12}^2=0$ with $s_{13}^2$ fixed at 0.1. 
Although we use the full probabilities including the 
matter effect, for 0.004 eV$^2$ this is not so important and one can 
understand the histograms from the vacuum oscillation probabilities. 
For fixed $\Delta_{23}$ as $s_{23}^2$ increases, $P_{\numu \numu}$ decreases, 
making the muon depletion higher. This is shown in the figure 
for two representative values of $\Delta_{23}$.
The electron events are not affected much by change of $\sbcsq$. 
The slight increase with $\sbcsq$ is due to increase of  
both $P_{\nue \nue}$ and $P_{\numu \nue}$. 
To understand the dependence on $\Delta_{23}$ we note that   
for $\sbcsq=0.2$, if one looks at the vacuum oscillation probabilities, 
$N_{\mu}/N_{\mu 0} \approx 1 - 0.65 S_{23}$.
For 0.004 eV$^2$
the contribution of $S_{23}$ is more resulting in a lower number of 
muon events. 
For the electron events however the behavior with $\Delta_{23}$ is 
opposite, with $N_e/N_{e0} = 0.82 +
0.12 S_{23}$. Thus with increasing $\Delta_{23}$ the number of electron 
events increase. Also note that 
since the contribution of $S_{23}$ comes with 
opposite sign 
the zenith-angle distribution for a fixed $\Delta_{23}$ is opposite for 
the muon and the electron events. 

In fig. 3c we show the histograms in the limit of $s_{13}^2$ = 0.0,
keeping $s_{12}^2$ as 0.1 and varying $\Delta_{23}$ and $s_{23}^2$. 
As $s_{23}^2$ increases all the relevant probabilities 
decrease
and therefore both $N_{\mu}/N_{\mu 0}$ and 
$N_{e}/N_{e 0}$ 
decrease giving less number of events for both. 
For this case the $S_{23}$ term comes with the same sign (negative) in both 
$N_{\mu}/N_{\mu 0}$ and $N_{e}/N_{e 0}$.
Therefore the depletion is more for higher $\Delta_{23}$  
for both muon and electron events.

Finally, the long dashed line in fig. 3d represent the
histograms for the best-fit value for 
two-generation $\numu-\nutau$ oscillations, for which $P_{\nue\nue}=1$.
The short dashed line gives the histograms for the
three-generation best-fit values. Both give comparable explanation 
for the zenith angle distribution of the data. The dotted line 
gives the event distribution for $\Delta_{23} = 10^{-5}$ eV$^2$. 
As discussed in section 3.2 even 
for such low value of $\Delta_{23}$, we find that due to the 
unique feature of the beyond OMSD neutrino mass spectrum, 
earth matter effects ensure that  
both the sub-GeV as well as the multi-GeV upward muon events are 
very well reproduced, as are the electron events. But since $\sabsq$ 
is high, the downward $\numu$ are depleted more than the data requires 
(eq. (6)). 

\subsection{Allowed parameter region}

In fig. 4a the solid lines give the 99\% C.L. allowed area 
from SK data in the 
$\Delta_{23}$-$s^2_{23}$ plane keeping the values of
$s^2_{13}$ and $s^2_{12}$ fixed in the 
allowed range from fig. 2 and Table 1. 
The first panel represents the two-generation $\numu - \nutau$ oscillation 
limit modulo the difference in the definition of the C.L. limit as the 
number of parameters are different. 
We have seen from the histograms in fig. 3a that raising $\sabsq$ 
results in electron excess and muon depletion. 
On the other hand 
increase in $\sacsq$ causes electron depletion. 
The above features are reflected in the shrinking 
and disappearance of the 
allowed regions in the first row and column. In the panels where both 
$\sabsq$ and $\sacsq$ are nonzero one may get allowed regions only when the 
electron depletion due to increasing $\sacsq$ is replenished by the 
increase in $\sabsq$.  

In fig. 4b we present the 99\% C.L.  
allowed areas 
in the  bilogarithmic $\tan^2\theta_{12} - \tan^2\theta_{13}$ plane for various
fixed values of the parameters $\Delta_{23}$ and $s^2_{23}$.
We use the $\log(\tan)$ representation which enlarges the allowed regions
at the corners and the clarity is enhanced.  
The four corners in this plot refer to the two-generation limits discussed
in section 3.
The extreme left corner ($\theta_{12} \rightarrow 0, \theta_{13}
\rightarrow 0$) correspond to 
the two generation $\nu_\mu - \nu_\tau$ oscillation limit.
As we move up increasing $\theta_{13}$,  
one has $\nu_e - \nu_\mu$ and $\nu_e - \nu_\tau$ mixing 
in addition and for $\sacsq \rightarrow 1 $ one
goes to the two generation 
$\nu_\mu - \nu_e$ oscillation region.
For the best-fit values of $\Delta_{23}$ and $\sbcsq$
if we take $\sabsq$ and $\sacsq$ to be 0 and $1$ respectively, 
then the $\chi^2_{min}$ is 66.92 which is therefore ruled out. 
Both the right hand corners in all the panels 
refer to pure $\nu_e - \nu_\tau$ 
oscillations and therefore there are no allowed regions in these zones.
For the panels in the first row, $\Delta_{23}=0.006$ eV$^2$ and  
the 2-generation $\numu-\nutau$ oscillation limit is 
just disallowed (this can also be seen in the first 
panel of fig. 4a). The small area allowed for the middle 
panel of first row (between the solid lines) 
is due to the fact that for non-zero 
$\sabsq$ and $\sacsq$ the electron events are better 
reproduced, while $\sbcsq=0.5$ takes care of the muon events. 
Hence for this case
slight mixture of $\numu-\nue$ and $\nue-\nutau$ oscillations
is favoured. 
This feature was also reflected in the fact that 
in the fig. 4a, the panel for $\sabsq=0.1$ and $\sacsq=0.3$ has 
more allowed range for $\Delta_{23}$ than 
the panel for the 2-generation $\numu-\nutau$ limit. 
 For the panels with $\Delta_{23}=0.002$ eV$^2$, both
the pure $\numu-\nutau$ limit as well as full three-generation 
oscillations, give good fit.
For the last two rows with $\Delta_{23}= 0.0007$ eV$^2$ and $0.0004$ eV$^2$ 
the matter effects are important in controlling the 
shape of the allowed regions. 
Infact the allowed region that one gets for $0.0004$ eV$^2$ and 
$s_{23}^2$ = 0.5 is the hallmark of the matter effect in this 
particular three-generation scheme. As can be seen from fig. 2a and 
Table 1,
if one does not include the matter effect, then there are no allowed 
regions below $\Delta_{23}$ = 0.0005 eV$^2$ for any arbitrary 
combination of the other three parameters. 
Even for the first and the last panels with $\Delta_{23}=0.0007$ eV$^2$, 
one gets allowed areas solely due to matter effects.

In fig. 4c the solid lines show the 99\% C.L. allowed regions 
from SK data in the 
$s^2_{23} - s^2_{12}$ plane for fixed values
of $\Delta_{23}$ and $s^2_{13}$.
In contrast to the previous figure, 
here (and in the next figure)  we use the $\sin-\sin$ representation 
because the allowed regions are around $\theta_{23} =\pi/4$ and this
region gets compressed in the $\log(\tan)-\log(\tan)$ representation. 
For explaining the various allowed regions we separate the figures in two sets
\begin{itemize}
\item For $s^2_{13}$ = 0.0, 
the four corners of the panels represent the no-oscillation limits  
inconsistent with the data. Also as discussed in section 3 
for $s_{23}^2$ = 0.0 or 1.0 one goes to the limit of 
pure $\nu_\mu-\nu_e$ conversions driven by
$\Delta_{LSND}$, which is not consistent with data. One obtains 
allowed regions only when 
$s^2_{23}$ is close to 0.5 with $s_{12}^2$ small, so that
$\numu-\nutau$ conversions are dominant.  
The allowed range of $\sabsq$ is controlled mainly by the electron 
excess as has been 
discussed before while the allowed range of $\sbcsq$ is determined 
mostly by the muon depletion.  

\item 
For $s_{13}^2 \neq 0$, the four corners represent
the two-generation $\nu_e - \nu_\tau$  oscillation 
limit discussed in section 3 and hence these corners are not allowed. 
For $s_{23}^2=0.0$ or 1.0 and $s_{12}^2 \neq $ 0 or 1 
one has $\Delta_{LSND}$ driven $\numu-\nue$ and $\numu-\nutau$ conversion and 
$\Delta_{ATM}$ driven $\nu_e-\nu_\tau$ conversions. 
This scenario is not allowed as it gives excess of electron events
and also fails to reproduce the correct zenith angle dependence.
For a fixed $\Delta_{23}$ as $s_{13}^2$ increases the
electron depletion increases 
which can be balanced by increasing $s_{12}^2$ which increases the number
of electron events. Hence for a fixed $\Delta_{23}$ 
the allowed regions shift towards higher $s_{12}^2$ values. 

As in fig. 4b the allowed area in the middle panel of the last row is 
due to the inclusion of the matter effect. 
\end{itemize}

In fig. 4d the solid contours refer to the  99\% C.L. allowed areas
from SK atmospheric neutrino data 
in the  $s^2_{13} - s^2_{23}$ plane for various values
of $\Delta_{23}$ and $s^2_{12}$. 

\begin{itemize}
\item For $s_{12}^2$ = 0.0 
the corners represent no oscillation limits.
In the limit $\sbcsq \rightarrow 0$ or 1, one gets $\nue-\nutau$ 
oscillation driven by $\Delta_{LSND}$ which is also not allowed. 
For $s_{13}^2$ = 0.0 and $\sbcsq \sim 0.5$ one has maximal two-flavour 
$\numu-\nutau$ oscillation limit which is therefore allowed (not 
allowed for $\Delta_{23} = 0.006$ eV$^2$ as discussed before).    
As $s_{13}^2$ increases the electron depletion 
becomes higher and that restricts 
higher $s_{13}^2$ values. 

\item For $s_{12}^2 \neq 0$, 
the four corners represent two-generation limits driven by
$\Delta_{LSND}$. This is the regime of average oscillations
and cannot explain the zenith angle dependence of the data.
For a fixed $\Delta_{23}$ the allowed region first expands and then  
shrinks in size and also shifts towards higher $s_{13}^2$ values as
$s_{12}^2$ increases just as in fig. 4c.

Matter effect is important for the last two rows and the increase 
in the allowed areas for the last two panels of $\Delta_{23}=0.0004$ 
eV$^2$ are typical signatures of matter effect. 

\end{itemize} 

In fig. 4e we present the allowed range in the $\Delta_{23}-\sbcsq$ 
plane with $\Delta_{23}$ in the $10^{-5} - 10^{-4}$ eV$^2$ range 
and $\sabsq$, $\sacsq$ fixed at 0.185 and 0.372 respectively. We get 
allowed regions in this range of small $\Delta_{23}$ and small 
mixing due to matter effects -- a feature unique to the mass spectrum 
considered in this paper.
 
\section{$\chi^2$ analysis of the SK + CHOOZ data}

The CHOOZ experiment can probe upto $10^{-3}$ eV$^2$ and hence it can be
important to cross-check the atmospheric neutrino results.
In particular a two-generation analysis shows that CHOOZ data disfavours
the $\nu_\mu - \nu_e$ solution to the atmospheric neutrino problem.
The general expression for the survival probability of the electron neutrino
in presence of three flavours is
\begin{equation}
P_{\nu_e \nu_e} = 1 - 4 U_{e1}^2 ( 1 - U_{e1}^2){\sin^2}
({\pi L /\lambda_{12})} - 4 U_{e2}^2 U_{e3}^2
{\sin^2}({\pi L/\lambda_{23}})
\label{pnue3}
\end{equation}
This is the most general expression without the one mass scale dominance
approximation.
We now minimize the $\chi^2$ defined as 
\begin{equation}
\chi^2 = \chi^2_{ATM} + \chi^2_{CHOOZ}
\label{comb}
\end{equation}
where we define
$\chi^2{_{CHOOZ}}$ as \cite{yasuda}
\begin{equation}
\chi^2_{CHOOZ} = \sum_{j=1,15}(\frac{x_{j} - y_{j}}{\Delta x_{j}})^2
\end{equation}
where $x_{j}$ are the experimental values, $y_{j}$ are the corresponding
theoretical predictions and the sum is over 15 energy bins of data of the
CHOOZ experiment \cite{chooz}. 
For the CHOOZ experiment the $\sin^2(\pi L/\lambda_{12})$ term does not 
always average out to 0.5 (for SK this term always averages to 0.5)
and one has to do the energy integration properly.
For our analysis we keep the $\Delta_{12}$ fixed at 0.5 eV$^2$ and do
a four parameter fit as in SK. 
The $\chi^2_{min}$ and the best-fit values of parameters that we get are
\begin{itemize}
\item
$\chi^2_{min}/d.o.f. = 42.22/51$, 
$\Delta_{23}$ = 0.0023 eV$^2$, $s^2_{23} = 0.5$, $s_{12}^2 = 0.0022$
and $s_{13}^2 = 0.0$. 
\end{itemize}
Thus the best-fit values shift towards the two-generation
limit when we include the CHOOZ result.  
This provides a very good fit to the data 
being allowed at 80.45\% C.L. 

The dotted lines in fig. 2 give 
the combined SK+CHOOZ $\Delta \chi^2 (= \chi^2 - \chi^2_{min})$ 
given by eq. (\ref{comb}), as a function of 
one of the parameters, keeping the other three unconstrained. 
We find that the CHOOZ data severely restricts the allowed ranges 
for the parameters $\sabsq$ and $\sacsq$ to values 
$\stackrel {<}{\sim} 0.047$, while $\Delta_{23}$ and 
$\sbcsq$ are left almost unaffected. 
Since CHOOZ is consistent with no oscillation one requires 
$P_{\nue\nue}$ close to 1. So the second and the third terms in eq. 
(\ref{pnue3}) should separately be very small. The second term 
implies $U_{e1}^2$ to be close to either 0 or 1. $U_{e1}^2$ close 
to zero implies either $\sabsq$ or $\sacsq$ close to 1 which 
is not consistent with SK. Therefore $U_{e1}^2$ is close to 1. 
Then from unitarity both $U_{e2}^2$ and $U_{e3}^2$ are close to 0 and 
so the third term goes to zero irrespective of the value of 
$\Delta_{23}$ and $\sbcsq$. Hence contrary to expectations, CHOOZ 
puts {\it almost} no restriction on the 
allowed values of $\sbcsq$ and $\Delta_{23}$, 
although $\Delta_{23} \sim 10^{-3}$ eV$^2$ -- in the regime in which 
CHOOZ is sensitive. On the other hand it puts severe constraints 
on the allowed values of $\sabsq$ and $\sacsq$ in order to 
suppress the average oscillations driven by $\Delta_{12}$.
Because of such low values of
$\sabsq$ and $\sacsq$ the matter effects for the atmospheric
neutrinos are not important and the additional allowed area with
low $\Delta_{23}$ and high $\sbcsq$ obtained in the SK analysis
due to matter effects are no longer allowed.
The 99\% C.L. regions allowed by 
a combined analysis of SK and CHOOZ data is shown by the dotted lines
in figs. 4a-d. 
It is seen that most of the regions allowed by the three-flavour analysis
of the SK data is ruled out when we include the CHOOZ result. 
None of the allowed regions shown in fig. 4a are allowed excepting the 
two-generation $\numu-\nutau$ oscillation 
limit because CHOOZ does not allow such high values of
either $s_{13}^2$ or $s_{12}^2$. Hence we present again 
in fig. 5 the allowed regions in the $\Delta_{23}-
\sbcsq$ plane for various fixed values of $\sabsq$ and 
$\sacsq$, determined from the dotted lines in fig. 2. 
The solid lines in fig. 5 give the 99\% C.L. area allowed
by the SK data while the dotted lines give the corresponding allowed 
region from the combined analysis of SK+CHOOZ.  
We find that for the combined analysis 
we get allowed regions in this plane only for much smaller values
of $s_{12}^2$ and $s_{13}^2$, which ensures that the electron events are 
neither less nor more than expectations.

\section{Combined allowed area from short baseline accelerator 
and reactor experiments}

As mentioned earlier the higher mass scale of this scenario 
can be explored in the accelerator based
neutrino oscillation search experiments. 
For the mass-pattern considered the most constraining 
accelerator experiments 
are LSND \cite{lsnd}, CDHSW \cite{cdhs}, 
E531 \cite{e531} and KARMEN \cite{karmen}.
Among these only LSND reported positive
evidence of oscillation. Other experiments are consistent with 
no-oscillation hypothesis. 
Also important in this mass range are the constraints from the reactor
experiment Bugey \cite{bugey}. 
The relevant probabilities are \cite{ska}
\begin{itemize}
\item Bugey
\begin{equation}
P_{\overline{\nu}_e\overline{\nu}_e} = 1 - 4c_{13}^2c_{12}^2
{\sin^2}({\pi L/\lambda_{12}})  + 
4c_{13}^4c_{12}^4 {\sin^2}({\pi L/\lambda_{12}})
\label{bugey} 
\end{equation}    
\item CDHSW
\begin{equation}
P_{\overline{\nu}_{\mu}\overline{\nu}_{\mu}} = 1 -
4 c_{12}^2 s_{12}^2 {\sin^2}({\pi L/\lambda_{12}}) 
\end{equation}
\item LSND and KARMEN
\begin{equation}
P_{\overline{\nu}_{\mu}\overline{\nu}_e} = 4 c_{12}^2 s_{12}^2
c_{13}^2 {\sin^2}({\pi L/\lambda_{12}})
\label{pnumul}
\end{equation}
\item{E531}
\begin{equation}
P_{\nu_{\mu} \nu_{\tau}} = 4 c_{12}^2 s_{12}^2 s_{13}^2 \sin^2({\pi
L/\lambda_{13}}) 
\label{pe531}
\end{equation}
\end{itemize}
We note that the probabilities are functions of one of the mass scales
and two mixing angles. Thus the one mass scale dominance approximation
applies. There are many analyses in the literature of the accelerator
and reactor data including LSND under this one mass scale dominance 
assumption \cite{ska,fogli95}.  
These analyses showed that when one considers the results from
the previous (prior to LSND) 
accelerator and reactor experiments there are three
allowed regions in the $\theta_{12} - \theta_{13}$ plane \cite{ska,fogli95}
\begin{itemize}
\item low $\theta_{12}$ - low $\theta_{13}$
\item low $\theta_{12}$ - high $\theta_{13}$
\item high $\theta_{12}$ - $\theta_{13}$ unconstrained
\end{itemize}
When the LSND result was combined with these results then only the
first and the third 
zones remained allowed in the mass range $0.5 \leq \Delta_{12} \leq 2$ eV$^2$.
In these earlier analyses of the accelerator and 
reactor data \cite{ska,fogli95}
E776 \cite{e776} was more constraining than KARMEN. But with the new 
data KARMEN2 gives stronger constraint than E776. Also the 
results from the KARMEN2 experiment now rule out most of the region 
allowed by the LSND experiment above 1 eV$^2$ \cite{karmen}. The LSND          
collaboration has also now reduced the value of the transition 
probability that they see \cite{newlsnd}.
We have repeated the analysis with the latest LSND and KARMEN results 
for one representative value of $\Delta_{12}=0.5$ eV$^2$ 
and present the allowed region in fig. 6.

The light-shaded area in fig. 6 shows the 90\% C.L. allowed area in the 
bilogarithmic $\tan^2\theta_{12}-\tan^2\theta_{13}$ 
plane from the observance of 
no-oscillation in all the other 
above mentioned accelerator and reactor experiments
except KARMEN2. The inclusion of the KARMEN2 results as well gives the 
90\% C.L. region shown by the area shaded by asterix. 
The 90\% allowed region by the LSND experiment is within the 
dashed lines. The KARMEN2 data severely restricts the LSND allowed regions.
The solid line shows the 90\% C.L. ($\chi^2 \leq \chi^2_{min} + 7.78$) 
region allowed by the combined 
$\chi^2$ analysis of the SK+CHOOZ data keeping $\Delta_{23}$ and $s_{23}^2$
at 0.002 eV$^2$ and 0.5 respectively. 
The combined SK  
atmospheric and the CHOOZ reactor data rule out 
the third zone (high $\theta_{12}$ with $\theta_{13}$
unconstrained ) allowed from LSND and other accelerator and reactor
experiments. 
Thus if one takes into account constraints from all experiments 
only a small region in the first zone (small $\theta_{12},\theta_{13}$) 
remains allowed. This common 
allowed region is shown as a dark-shaded area in the fig. 6. 
As evident from the expression of the probabilities for the 
accelerator and reactor experiments the 
combined allowed area of all the accelerator reactor 
experiments remains the 
same irrespective of the value of $\Delta_{23}$ and $\sbcsq$.  
Even though the combined area in fig. 6 
shows that in the 
first zone (small $\theta_{12},\theta_{13}$), SK+CHOOZ data 
allows more area in the $\theta_{12}-\theta_{13}$ plane for 
$\Delta_{23} = 0.002$ eV$^2$ and $\sbcsq =0.5$, 
from fig. 4b we see that for some other combinations of 
$\Delta_{23}$ and $\sbcsq$ one does not find any allowed zones 
from the SK+CHOOZ analysis, even at 
99\% C.L.. For those sets of values of $\Delta_{23}$ and $\sbcsq$ 
the SK+CHOOZ analysis is more restrictive than the LSND and other 
accelerator reactor data. 

\section{Implications}

From our analysis of the SK atmospheric data the explicit 
form for the $3\times3$ mixing matrix $U$ at the best-fit values of parameters
is 

\begin{equation}
U = {\pmatrix {0.95 & -0.039 & 0.31 \cr
-0.2 & 0.686 & 0.7 \cr
-0.24 & -0.727 & 0.644 \cr}}
\end{equation}

From the combined SK+CHOOZ analysis the mixing matrix at the best-fit 
values of the parameters is 

\begin{equation}
U = {\pmatrix {0.999  & 0.033 & 0.033 \cr
-0.047 & 0.706 & 0.706 \cr
-0.0 & -0.707 & 0.707 \cr}}
\end{equation}

From the combined allowed area of fig. 6 the mixing matrix at 
$\Delta_{12}$ = 0.5 eV$^2$, $\Delta_{23} = 0.0028$ eV$^2$,
$s_{12}^2 = 0.005$, $s_{13}^2$ = 0.001 and
$s_{23}^2$ = 0.5, is 
\begin{equation}
U = {\pmatrix {0.997 & 0.028 & 0.072 \cr
-0.071 & 0.705 & 0.705 \cr
-0.032 & -0.708 & 0.705 \cr}}
\end{equation}

Thus the allowed scenario corresponds to 
the one where $\langle \nu_{1}|\nu_{e} \rangle$ is close to 1 
while the states $\nu_{2}$ and $\nu_{3}$ are combinations of
nearly maximally mixed $\nu_{\mu}$ and $\nu_{\tau}$ 
\footnote{ Thus this scenario is the same as the one termed 3a 
in Table VI in the pre-SK analysis
of \cite{flms97}. In their notation the states 2 and 3 were 1 and 2. 
It was disfavoured from solar neutrino results.}.  

Long baseline (LBL) experiments
can be useful to confirm if the atmospheric neutrino anomaly is
indeed due to neutrino oscillations, using well monitored 
accelerator neutrino beams.  
Some of the important LBL experiments are 
K2K\footnote{K2K has already presented some preliminary 
results.} (KEK to SK, L $\approx$ 250 km)\cite{k2k},
MINOS (Fermilab to Soudan, L $\approx$ 730 km ) \cite{minos} 
and the proposed CERN to Gran Sasso experiments (L $\approx$ 730 km)
\cite{cgs}. 
In this section we explore the sensitivity of the LBL experiment 
K2K in probing the parameter spaces allowed by the SK+CHOOZ and other 
accelerator and reactor experiments including LSND.
K2K will look for $\nu_\mu$ 
disappearance as well as $\nu_e$ appearance.
In fig. 7 we show the regions in the $\Delta_{23} - \sbcsq$ 
plane that can be probed by K2K using their 
projected sensitivity from \cite{k2k}.
The top left panel is for the two-generation $\nu_\mu - \nu_\tau$ limit. 
The other panels are for different fixed values of $\sabsq$ and $\sacsq$ 
while $\Delta_{12}$ is fixed at 0.5 eV$^2$. 
For LBL experiments the term containing 
$\Delta_{12}$ averages to 0.5 as in the atmospheric case. 
The solid lines in the panels show the region that can be probed by 
K2K using the $\numu$ disappearance channel while the dotted lines 
give the 90\% C.L. contours allowed by SK+CHOOZ. One finds that for  
for $\Delta_{23} \geq 2 \times 10^{-3}$ eV$^2$, the whole region allowed
by SK+CHOOZ can be probed by the $\nu_\mu$ disappearance channel in 
K2K. The dashed lines show the 90\% C.L. area that K2K can probe 
by the $\nue$ appearance mode. 
As $s_{12}^2$ increases the constraint from the $P_{\numu \nue}$ channel 
becomes important as is seen in the top right panel of fig. 7.   
However such high values of $s_{12}^2$, although allowed by SK+CHOOZ, is
not favoured when one combines LSND and other accelerator and reactor results. 
For lower $s_{12}^2$ values allowed by all the accelerator, reactor and
SK atmospheric neutrino experiment the projected sensitivity in the
$\numu-\nue$ channel of K2K is not enough to probe the allowed regions
in the $\Delta_{23} - \sbcsq$ plane as is shown by the absence of the 
dashed curves in the lower panels. 

In fig. 8 we show the regions in the bilogarithmic $\tan^2 \theta_{12} - 
\tan^2 \theta_{13}$ plane which can be probed by K2K.
For drawing these curves we fix $\Delta_{23}=0.002$ eV$^2$,  
$s_{23}^2=0.5$ and $\Delta_{12}=0.5$ eV$^2$. 
Shown is the area that 
can be explored by the $\numu - \numu$ (left of the solid line)
and $\numu - \nue$ (hatched area) 
channels in K2K at 90\% C.L.. 
The light-shaded area is allowed by SK+CHOOZ
and the dark shaded area is allowed by the combination of all the  
accelerator, reactor and SK atmospheric neutrino data at 90\% C.L.. 
It is clear from the figure that even though the 
sensitivity of the $\nue$ appearance channel is not 
enough, K2K can still probe the 
combined allowed region in the $\theta_{12}-\theta_{13}$ plane 
from $\numu$ disappearance.

The projected sensitivities of MINOS and the CERN to ICARUS proposals 
are lower than K2K and it will be interesting to check if one can
probe the regions allowed in this picture better in these experiments. 
However since in our case the OMSD approximation is not applicable 
one has to do the energy averaging properly 
to get the corresponding contours in the three-generation parameters space, 
and one cannot merely scale the allowed regions 
from the two-generation plots. For K2K we could use the
fig. 5 of \cite{k2k} to circumvent this problem. However since the 
analogous information for MINOS and CERN-Gran Sasso proposals is not 
available to us we cannot check this explicitly. 

An important question in this context is whether one can 
distinguish between the OMSD three generation and this mass scheme. 
In both pictures the SK atmospheric neutrino data can be explained
by the dominant $\nu_\mu-\nu_\tau$ oscillations  mixed with little amount of 
$\nu_e - \nu_\mu (\nu_\tau)$  transition.
However the mixing matrix $U$ is 
different. 
A distinction                               
can be done if one can measure the mixing angles very accurately.

What is the prospect in LBL experiments to distinguish between these
pictures?
We give below a very preliminary and qualitative  discussion on this.
If we take $s_{12}^2$ = 0.02, $s_{13}^2$ = 0.02 and $s_{23}^2$ = 0.5,  
$P_{\numu \nue}$ would be  (0.038 + 0.0004
$\langle S_{23} \rangle$).
As the second term is negligible one has average oscillations. 
This is different from the OMSD limit where $P_{\numu \nue} = 4 U_{\mu3}^2 
U_{e3}^2 S_{23}$ is energy dependent. 
If one combines the other accelerator and reactor 
experiments including LSND then the allowed values of   
of $s_{12}^2$ and $s_{13}^2$ are even less 
and choosing $s_{12}^2$ = 0.005, $s_{13}^2$ = 0.001 and $s_{23}^2$ = 
0.5 we get $P_{\nue \numu} = 0.01 - 0.004  \langle S_{23} \rangle$. 
Here also the term involving $\langle S_{23} \rangle$  
is one order of magnitude smaller and the oscillations will be averaged. 
Thus this channel has different predictions for the OMSD limit 
and beyond the OMSD limit.

\section{Discussions and Conclusions}

In this paper we have done a detailed $\chi^2$ analysis of the SK
atmospheric
neutrino data going beyond the OMSD approximation. 
The mass spectrum chosen is such that $\Delta_{12} = \Delta_{13} \sim$
eV$^2$ to explain the LSND data and $\Delta_{23}$ is
in the range suitable for the atmospheric neutrino problem. 
We study in details the implications of the earth matter 
effects and bring out the essential differences of our mass pattern 
with the OMSD scenario and the two-generation limits.

We first examine in detail 
what are the constraints obtained from only SK data considering
its overwhelming statistics. 
The allowed regions include
\begin{itemize}
\item the two-generation $\nu_\mu - \nu_\tau$ limit (both $s_{12}^2$ and 
$s_{13}^2$ zero) 
\item regions where either $s_{12}^2$ or $s_{13}^2$ is zero;
in this limit the probabilities are functions in general of two
mixing angles and two mass scales.
\item  
the three-generation 
regions with all three mixing angles non-zero and the probabilities 
governed by both mass scales.

The last two cases correspond to dominant $\numu-\nutau$ oscillation 
with small admixture of $\numu-\nue$ and $\nue-\nutau$ oscillation. 
\item
regions with very low $\Delta_{23}$ ($<10^{-4}$ eV$^2$) 
and $\sbcsq$ close to 1, for which 
the earth matter effects enhance the oscillations of the upward 
neutrinos and cause an up-down flux asymmetry. This region is 
peculiar to the mass spectrum considered by us and is 
absent in the two-generation and the OMSD pictures. 
\end{itemize} 
We present the zenith angle distributions of the events in these cases.
With the inclusion of the CHOOZ result the allowed ranges of the mixing angles
$s_{12}^2$ and $s_{13}^2$ is constrained more ($\stackrel{<}{\sim} 0.047$),  
however the allowed ranges of
$\Delta_{23}$ and $s_{23}^2$ do not change much (see fig. 2) except 
that the low $\Delta_{23}$ region allowed by SK due to matter effects 
is now disallowed. 
The inclusion of the constraints from LSND and other accelerator and
reactor experiments may restrict the allowed area in the 
$\theta_{12}-\theta_{13}$ plane for certain values of $\Delta_{23}$ and 
$\sbcsq$, but for some other combinations of $\Delta_{23}$ and 
$\sbcsq$, SK+CHOOZ turns out to be more constraining. 
We have included the latest results from LSND and KARMEN2 in our 
analysis.

In order to explain the solar neutrino problem in this picture one has to add 
an extra light sterlie neutrino. 
With the new LSND results the allowed 4 neutrino scenarios are 
\begin{itemize}
\item
the (2+2) picture where two degenerate mass states
are separated by the LSND gap \cite{sg,grimus,garcia,barger}.
\item
the (3+1) scheme with three neutrino states closely degenerate in mass and
the fourth one separated from these by the LSND gap \cite{barger,giunti4}.
In \cite{barger} the separated state is predominantly a sterile 
state. In \cite{giunti4}, on the other hand, the state sepaerated by the LSND 
gap has a very small sterile component.  
\end{itemize} 
The extension of our scenario to the 2+2 picture is 
straight forward. One has to add an extra sterile state 4 
close to the state 1 such that $\Delta_{14}$ is in the solar range. 
Then we have two almost decoupled two-generation 
pictures in which the atmospheric neutrino problem is 
mainly due to $\nu_\mu -\nu_\tau$ oscillation and the solar neutrino problem 
is explained by $\nu_e -\nu_s$ oscillation. The SMA MSW solution for two
-generation $\nu_e-\nu_s$  picture is allowed at $\sim$ 15\% C.L.
\cite{mswdk}. A detailed global fit of solar and atmospheric neutrino data 
under this picture would tell us how much this will change due to the 
small admixture with the other generations. 
If on the other hand we assume the 4th state to be close to the 2nd and 
the third state then we will have a (3+1) picture where the 
1 state, separated by the LSND gap,
is predominantly $\nu_e$. This picture will have difficulties in 
solving the solar neutrino problem as because of the CHOOZ constraints 
$U_{ei}$ (i=2,3,4) are small so that 
$P_{\nue \nue}$ $\approx$ 1 indicating very small suppression 
of the solar neutrino flux.   

To conclude, one can get allowed regions from the SK atmospheric 
neutrino data 
where both the mass scales and all the three mixing angles are 
relevant. The beyond one mass scale dominance spectrum considered 
in this paper allows new regions in the low mass -- low mixing 
regime due to the earth matter effects. 
With the inclusion of the CHOOZ, LSND and other accelerator 
reactor results, the allowed regions are constrained severely. 
It is, in principle, possible to 
get some signatures in the LBL experiments to 
distinguish this picture from the OMSD limit. 

\vskip 12mm

\noindent
{\small The authors wish to thank  Kenji Kaneyuki for sending them the  
1144 days SK atmospheric data and the detection efficiencies. 
They also wish to thank E. Lisi, N. Fornengo 
and S. Uma Sankar 
for useful correspondences.}

\newpage
\begin{center}
{\bf Figure Captions}
\end{center}

\noindent
Fig. 1: The two possible neutrino mass spectra in a three generation
scheme. 
\vskip 6mm

\noindent
Fig. 2: The variation  of $\Delta \chi^2 = \chi^2 - \chi^2_{min}$ 
with one of the parameters keeping the other three unconstrained. 
The solid (dashed) line corresponds to 
only SK data when matter effects are included (excluded) 
while the dotted curve gives the same for 
SK+CHOOZ. The dashed-dotted line shows the 99\% C.L. limit for 4 parameters. 
\vskip 6mm

\noindent
Fig. 3a: The zenith angle distribution of the lepton events 
with $\Delta_{23}=0.002$ eV$^2$ and $\sbcsq$ 
= 0.5 for various combinations of 
$\sabsq$ and $\sacsq$. 
$N$ is the number of events as given by eq. (1)
and $N_0$ is the corresponding number with survival probability 1.
The panels labelled $SG_{\alpha}$ and $MG_{\alpha}$ ($\alpha$ can be e or 
$\mu$) give the 
histograms for the sub-GeV and multi-GeV $\alpha$-events respectively.
Also shown are the SK experimental data points with $\pm$ 1$\sigma$ error
bars. 
\vskip 6mm

\noindent
Fig. 3b: Same as in fig. 3a 
for fixed $\sabsq=0.1$ and $\sacsq=0.0$ 
varying $\Delta_{23}$ and $\sbcsq$. 
\vskip 6mm

\noindent
Fig. 3c: Same as in fig. 3a  
fixing $\sabsq=0.0$ and $\sacsq=0.1$ for different 
$\Delta_{23}$ and $\sbcsq$ values.
\vskip 6mm

\noindent
Fig. 3d: The long-dashed (short-dashed) line gives the 
zenith angle distribution of the lepton events for the 
best-fit cases of the two-generation (three-generation) 
oscillation solutions for SK. The dotted line gives 
the corresponding distribution for $\Delta_{23}=10^{-5}$ eV$^2$, 
$\sabsq=0.2$, $\sacsq=0.4$ and $\sbcsq=1.0$. 
\vskip 6mm

\noindent
Fig. 4a: The allowed parameter regions in the $\Delta_{23} - 
\sbcsq$ plane for various fixed values of $\sabsq$ and 
$\sacsq$, shown at the top of each panel. The solid lines 
corresponds to the 99\% C.L. contours from the SK data alone, while 
the dotted line 
gives the 99\% contour from the 
combined analysis of the SK+CHOOZ data.
\vskip 6mm

\noindent
Fig. 4b: Same as 4a but in the bilogarithmic $\tan^2\theta_{12}-
\tan^2\theta_{13}$ plane for 
fixed values of $\Delta_{23}$ and $s_{23}^2$.
\vskip 6mm

\noindent
Fig. 4c: Same as 4a but in the $\sabsq-\sbcsq$ plane for fixed values
of $\sacsq$ and $\Delta_{23}$.
\vskip 6mm

\noindent
Fig. 4d: Same as 4a but in the $\sacsq-\sbcsq$ plane for various fixed
values of $\sabsq$ and $\Delta_{23}$.
\vskip 6mm

\noindent
Fig. 4e: The allowed parameter space in the $\Delta_{23}-\sbcsq$ plane 
with $\Delta_{23}$ in the range $10^{-5}-10^{-4}$ eV$^2$ and with 
fixed values of $\sabsq=0.185$ and $\sacsq=0.372$.

\noindent
Fig. 5: Same as 4a but for smaller values of $\sabsq$ and 
$\sacsq$, chosen from the range determined by the 
SK+CHOOZ dashed line in fig. 3.
\vskip 6mm

\noindent
Fig. 6: The area between the dashed lines is the 90\% C.L. region 
allowed by LSND while the 
light shaded zone gives the 90\% C.L. allowed region 
from the non-observance of neutrino oscillation in the other 
short baseline accelerator and reactor experiments except KARMEN2.  
The corresponding area which includes KARMEN2 as well is 
shown by the region shaded by asterix. 
The 90\% C.L. allowed region from 
SK+CHOOZ analysis is within the dotted line. 
The dark shaded area corresponds to the combined allowed region.  
\vskip 6mm

\noindent
Fig. 7: 90\% C.L. regions in the $\Delta_{23}-\sbcsq$ plane 
that can be explored 
by the $\numu-\numu$ (solid line) and $\numu-\nue$ (dashed line) 
oscillation channels in the K2K experiment. The area inside the dotted line 
shows the 90\% C.L. region allowed by SK+CHOOZ. The curves are presented 
for fixed values of $\sabsq$ and $\sacsq$ with $\Delta_{12}=0.5$ eV$^2$.
\vskip 6mm

\noindent
Fig. 8: Sensitivity of the K2K experiment in the $\tan^2\theta_{12}-
\tan^2\theta_{13}$ plane for $\Delta_{23}=0.002$ eV$^2$, 
$s_{23}^2=0.5$ and $\Delta_{12}=0.5$ eV$^2$. 
The area that 
can be explored by the $\numu - \numu$ (left of solid line)  
and $\numu - \nue$ (hatched area)  
channels in K2K at 90\% C.L. is shown. 
The light-shaded area is allowed by SK+CHOOZ
and the dark-shaded region is the combined area allowed by all 
accelerator and reactor data at 90\% C.L..



\begin{figure}[p]
\epsfxsize 16 cm
\epsfysize 17 cm
\epsfbox[25 151 585 704]{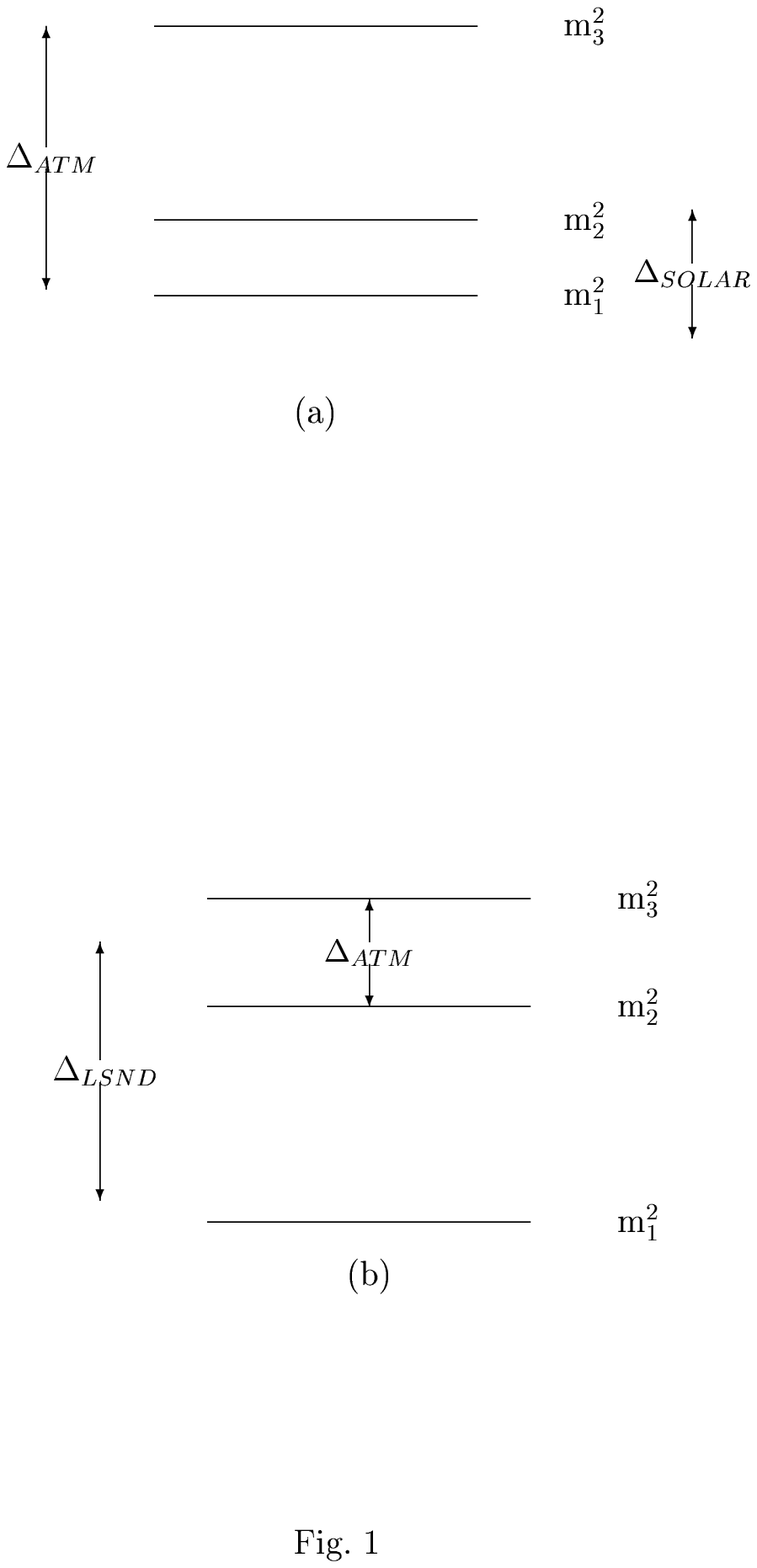}
\end{figure}

\begin{figure}[p]
\epsfxsize 16 cm
\epsfysize 17 cm
\epsfbox[25 151 585 704]{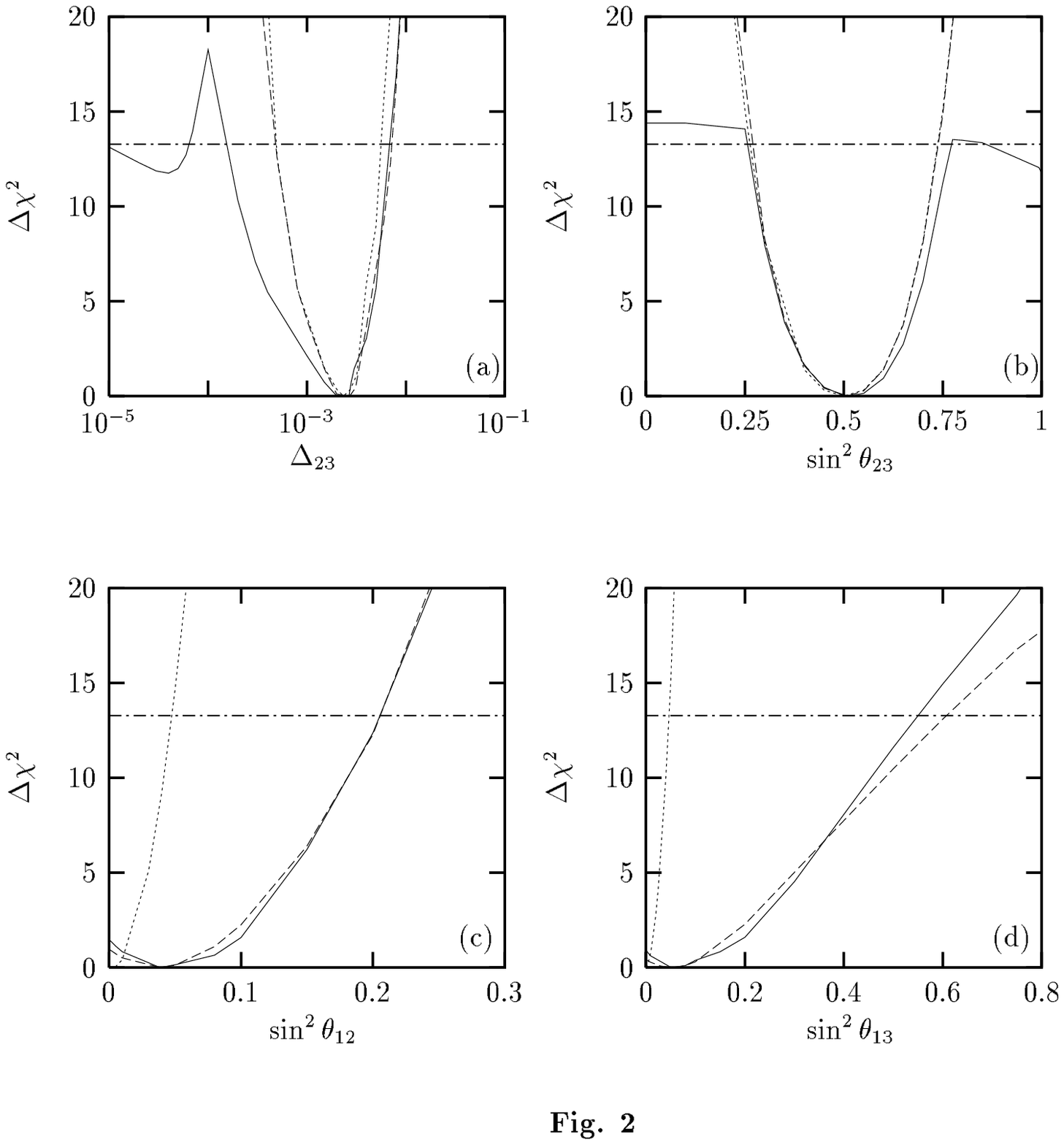}
\end{figure}
 
\begin{figure}[p]
\epsfxsize 16 cm
\epsfysize 17 cm
\epsfbox[25 151 585 704]{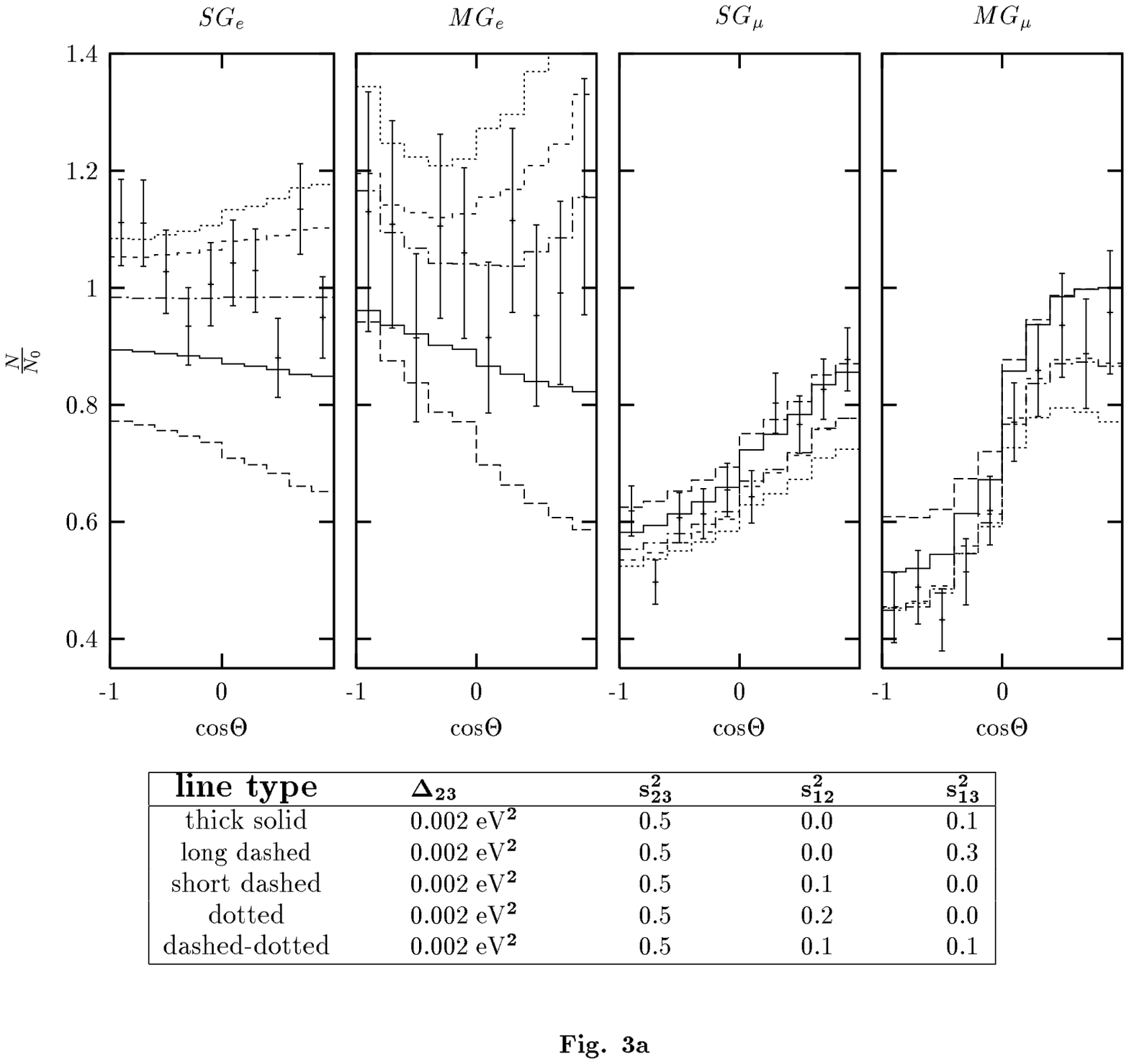}
\end{figure}

\begin{figure}[p]
\epsfxsize 16 cm
\epsfysize 17 cm
\epsfbox[25 151 585 704]{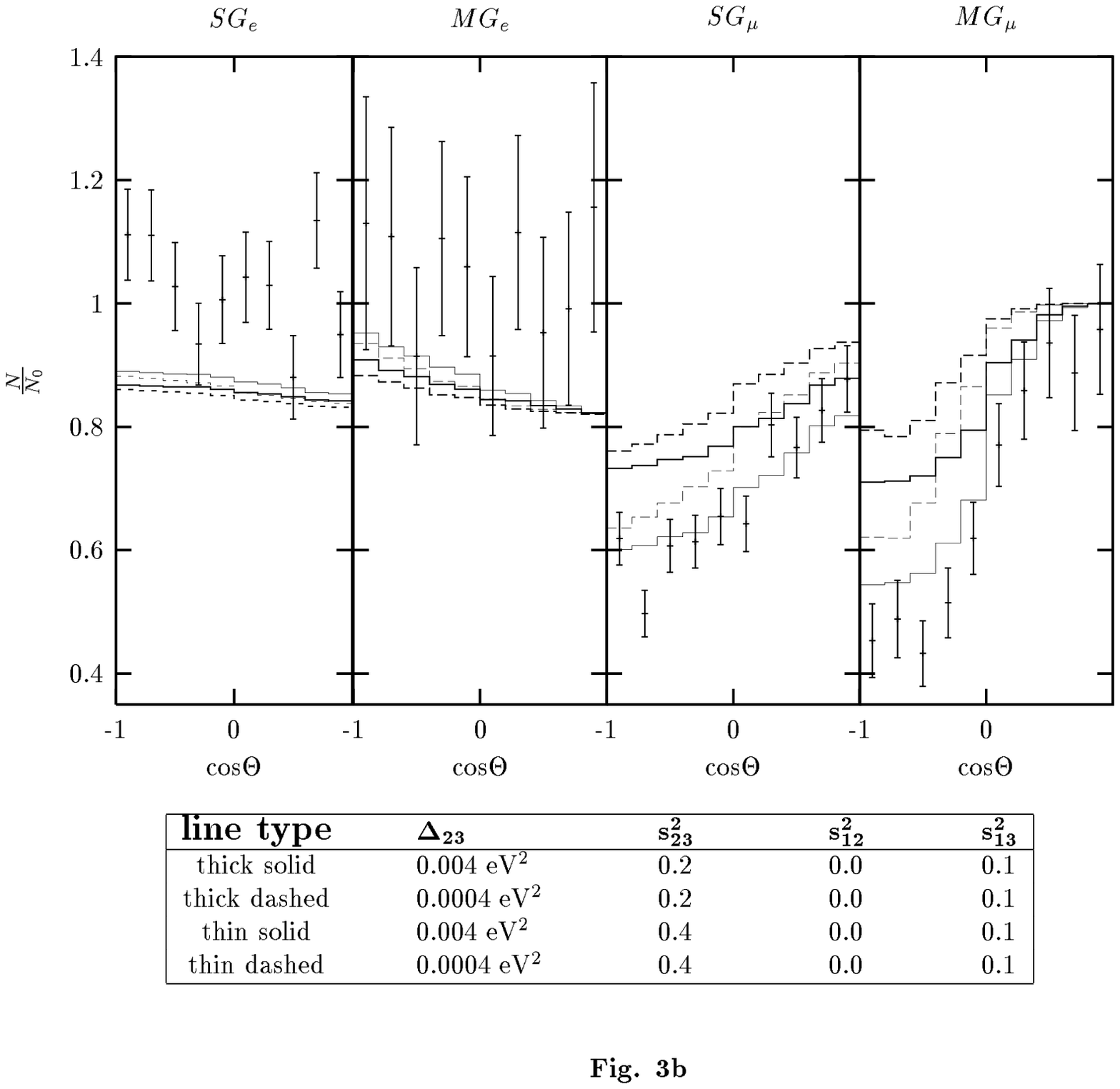}
\end{figure}

\begin{figure}[p]
\epsfxsize 16 cm
\epsfysize 17 cm
\epsfbox[25 151 585 704]{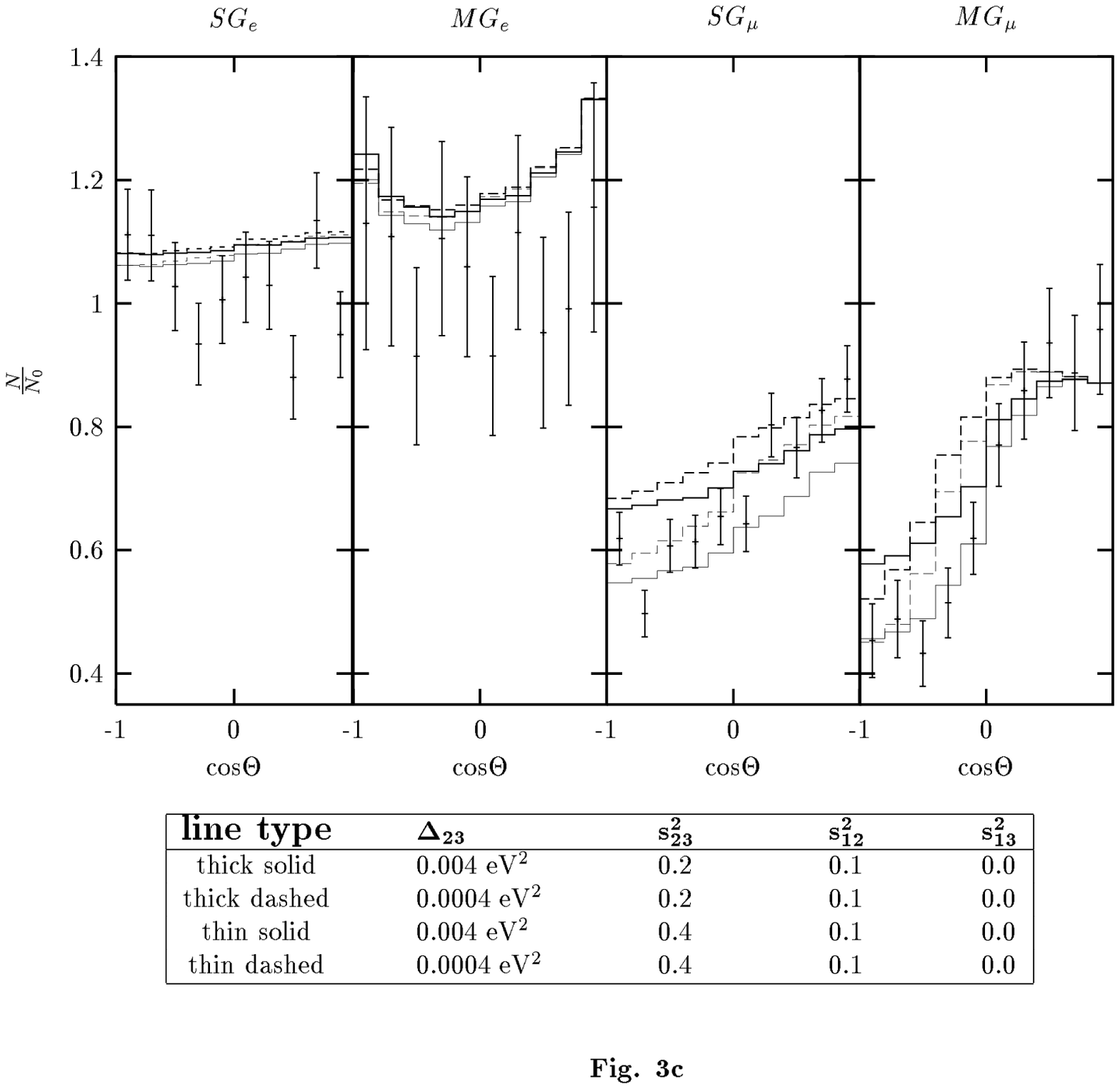}
\end{figure}

\begin{figure}[p]
\epsfxsize 16 cm
\epsfysize 17 cm
\epsfbox[25 151 585 704]{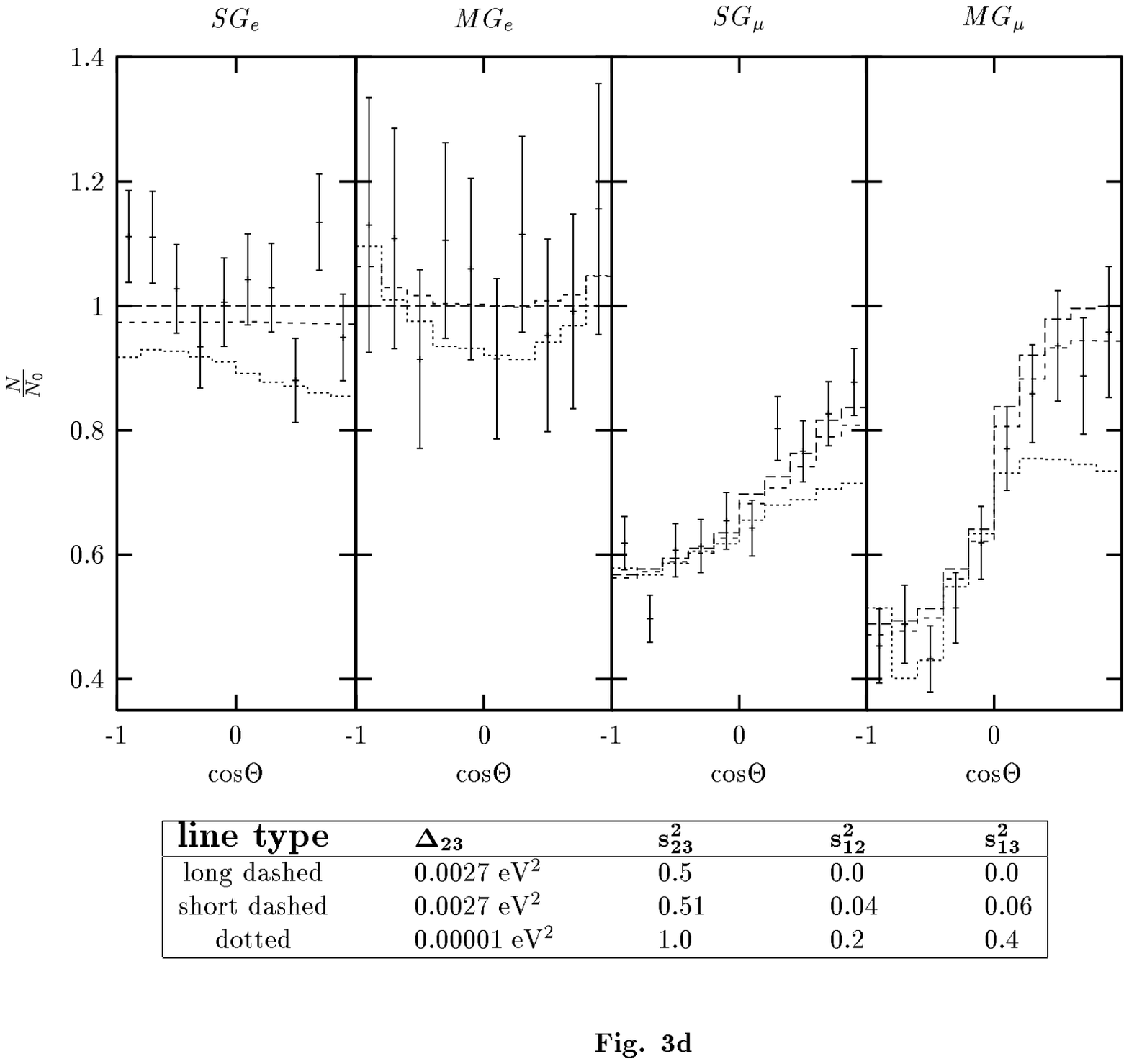}
\end{figure}

\begin{figure}[p]
\epsfxsize 16 cm
\epsfysize 17 cm
\epsfbox[25 151 585 704]{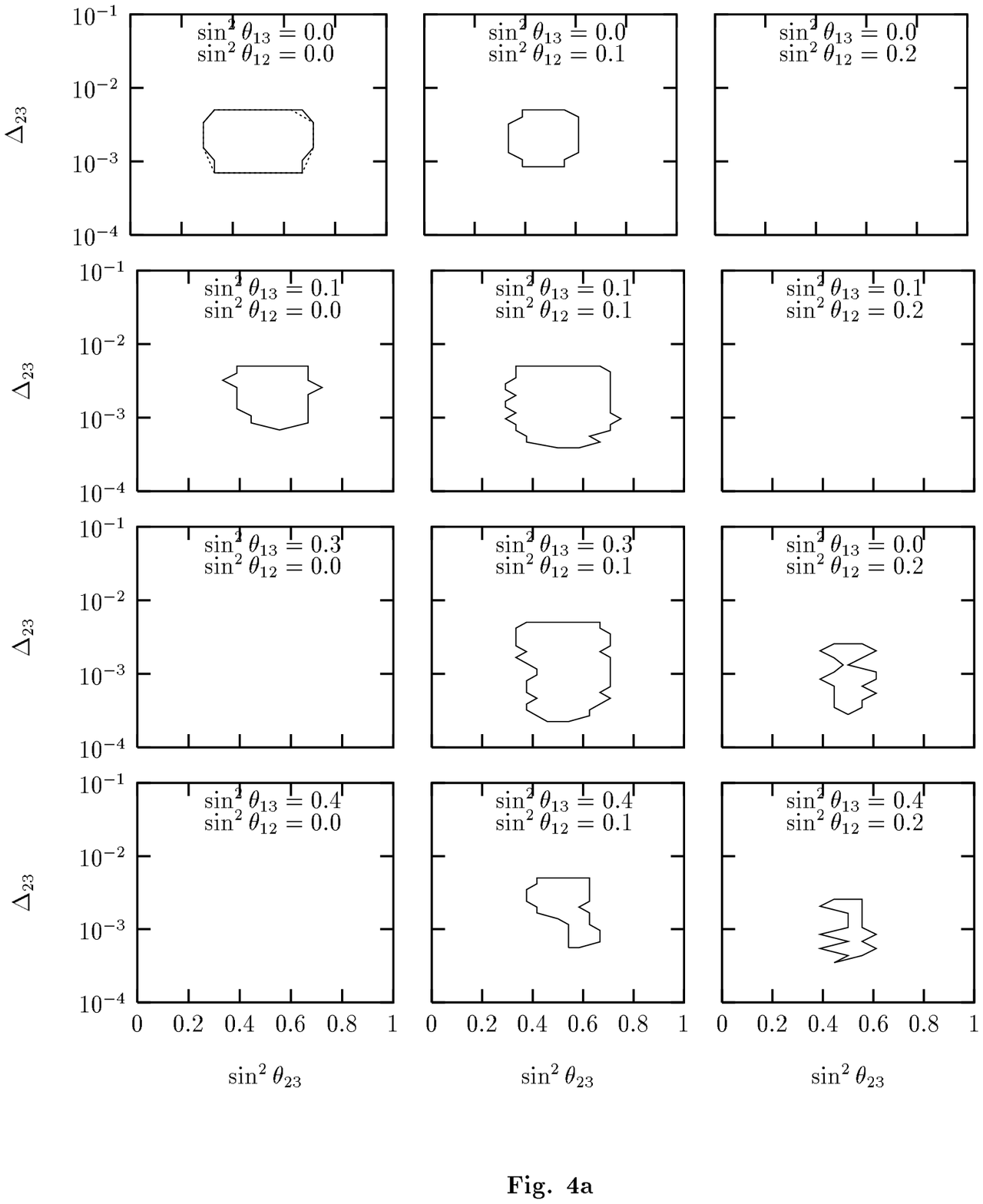}
\end{figure}

\begin{figure}[p]
\epsfxsize 16 cm
\epsfysize 17 cm
\epsfbox[25 151 585 704]{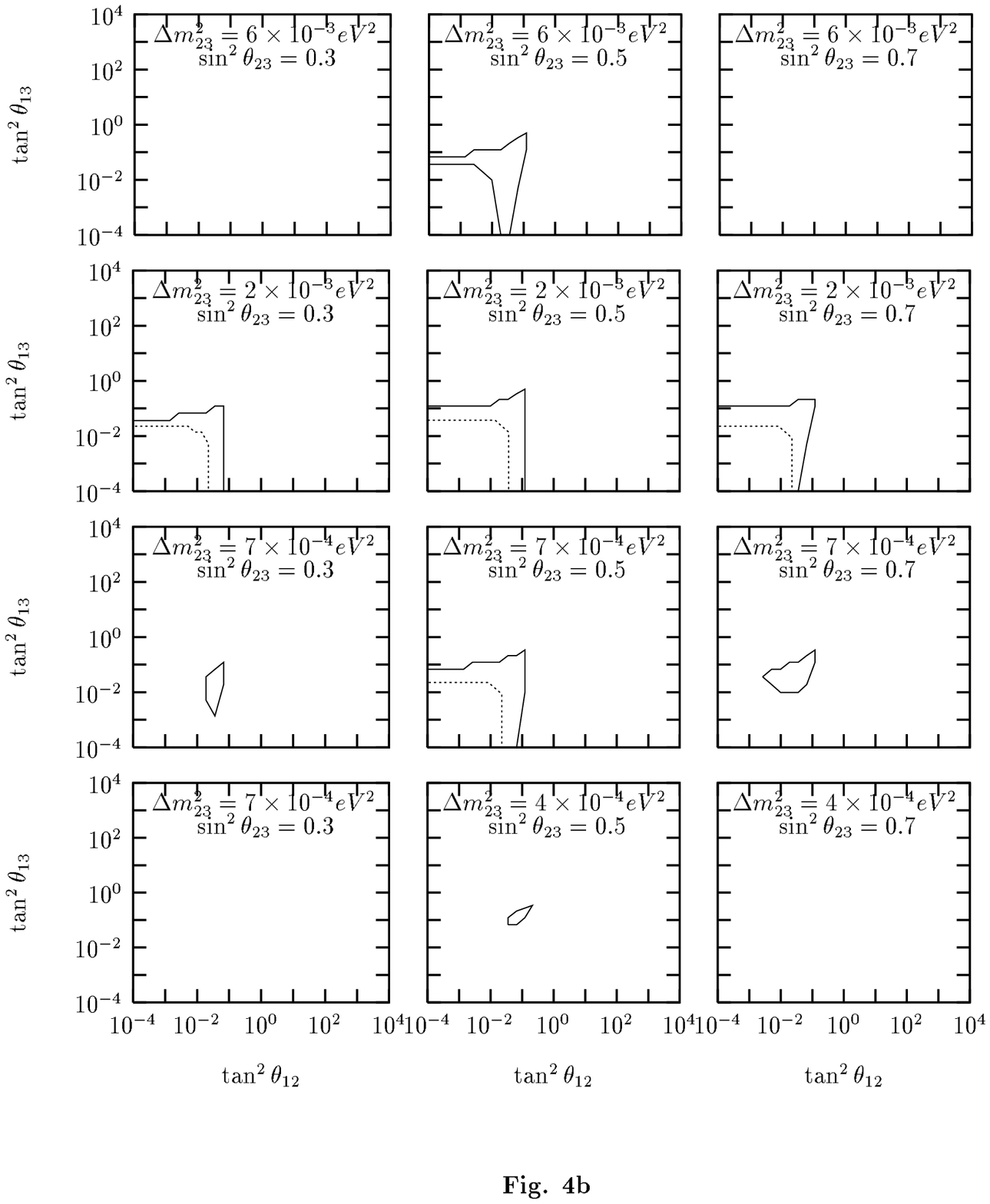}
\end{figure}

\begin{figure}[p]
\epsfxsize 16 cm
\epsfysize 17 cm
\epsfbox[25 151 585 704]{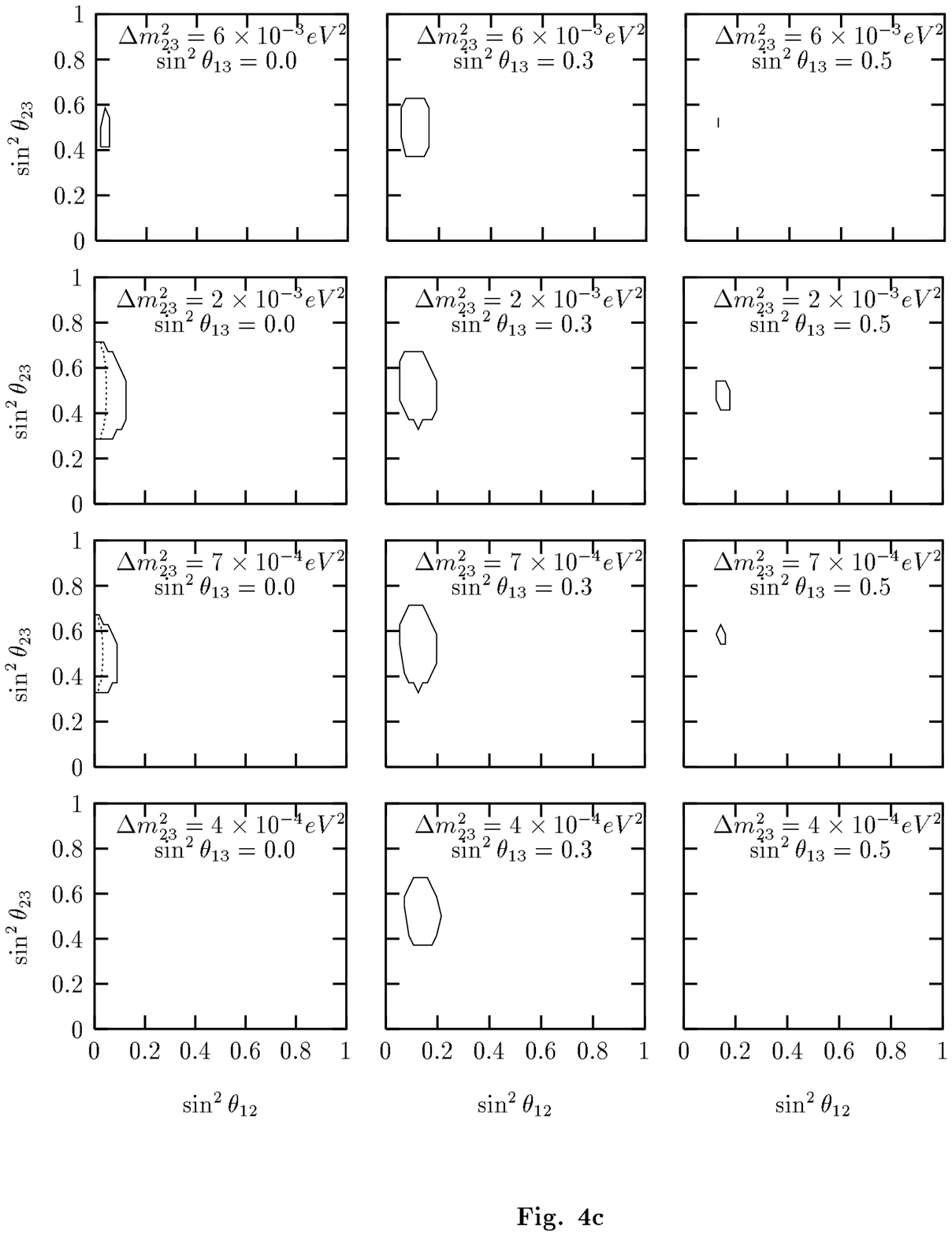}
\end{figure}

\begin{figure}[p]
\epsfxsize 16 cm
\epsfysize 17 cm
\epsfbox[25 151 585 704]{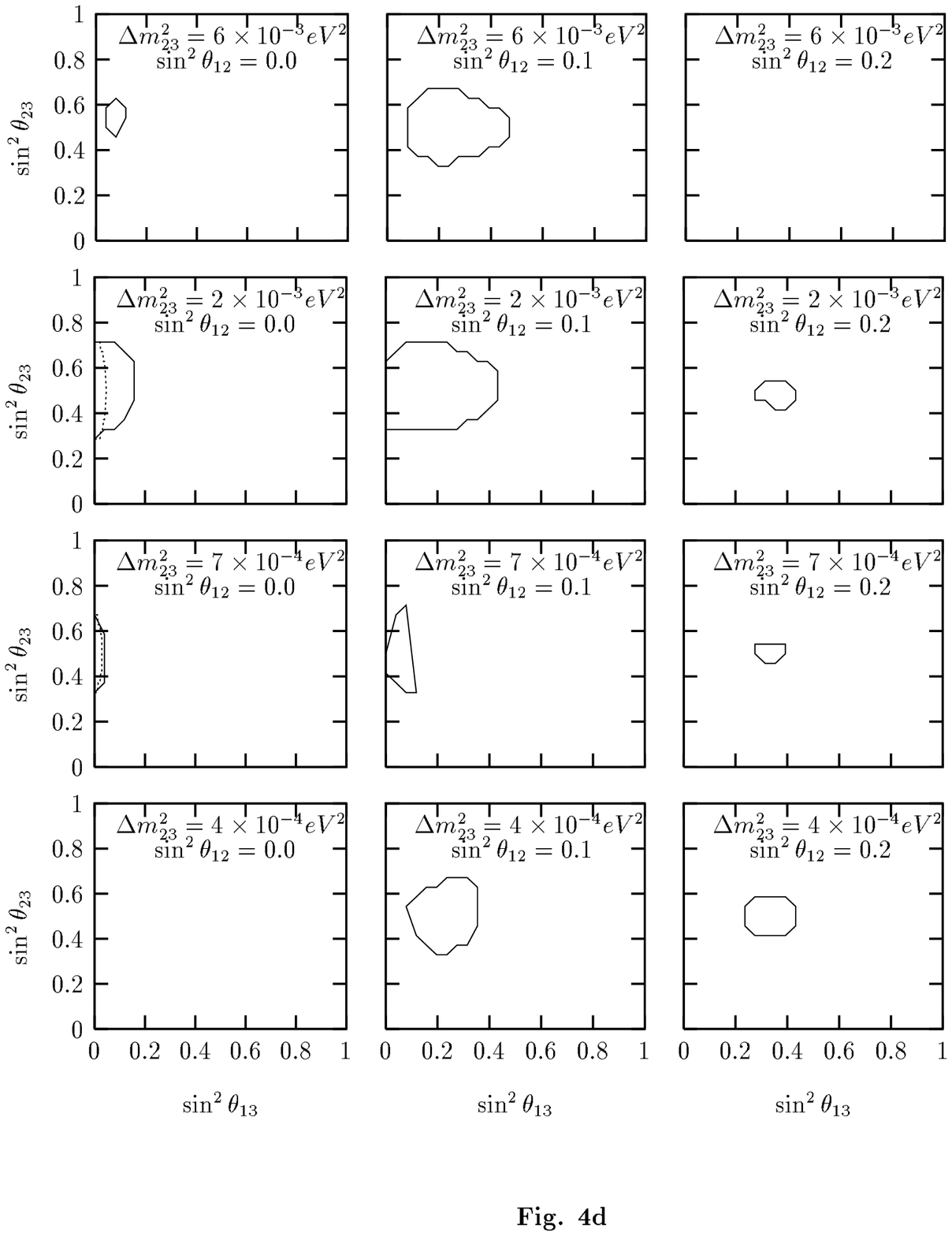}
\end{figure}

\begin{figure}[p]
\epsfxsize 16 cm
\epsfysize 17 cm
\epsfbox[25 151 585 704]{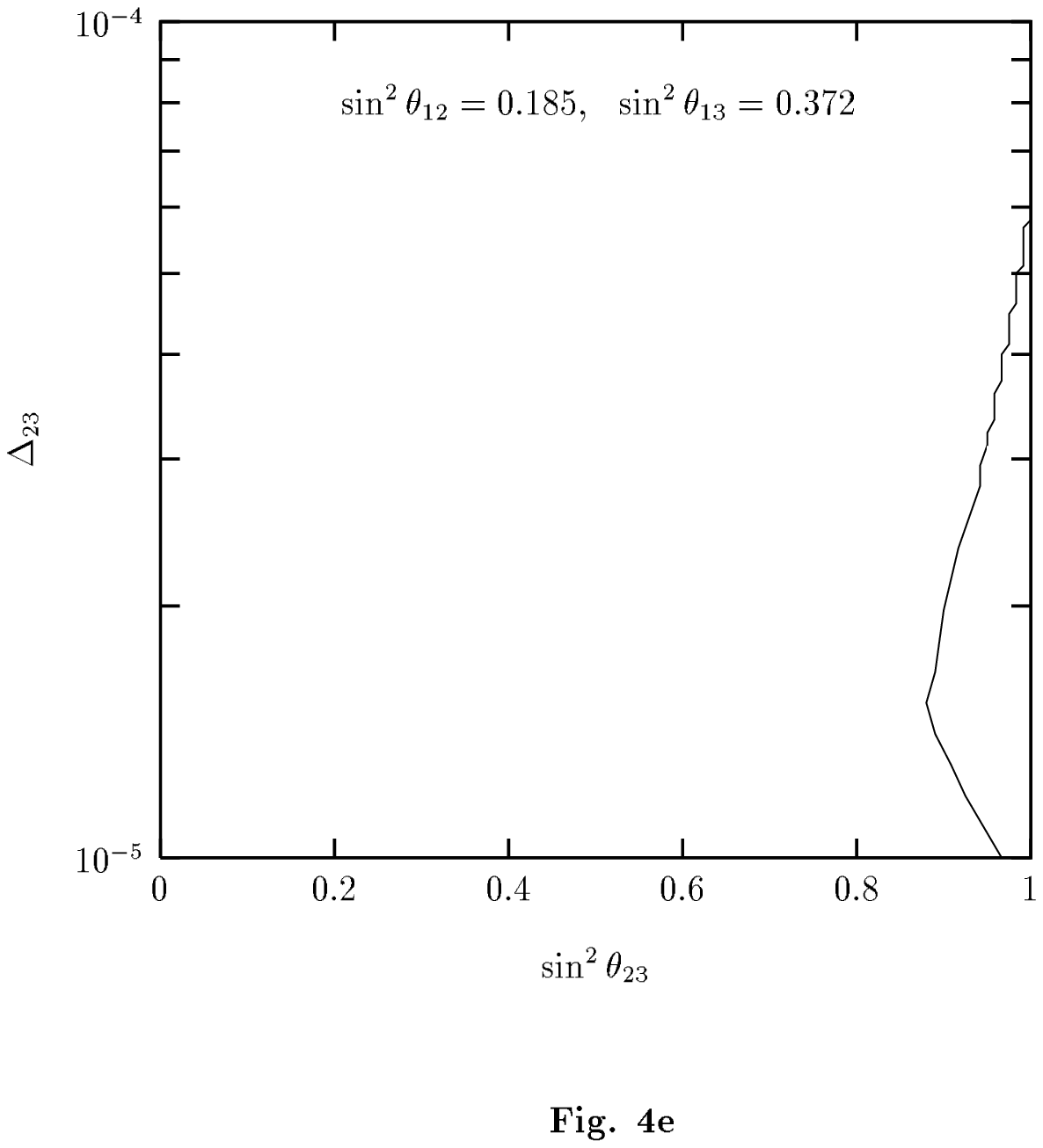}
\end{figure}

\begin{figure}[p]
\epsfxsize 16 cm
\epsfysize 17 cm
\epsfbox[25 151 585 704]{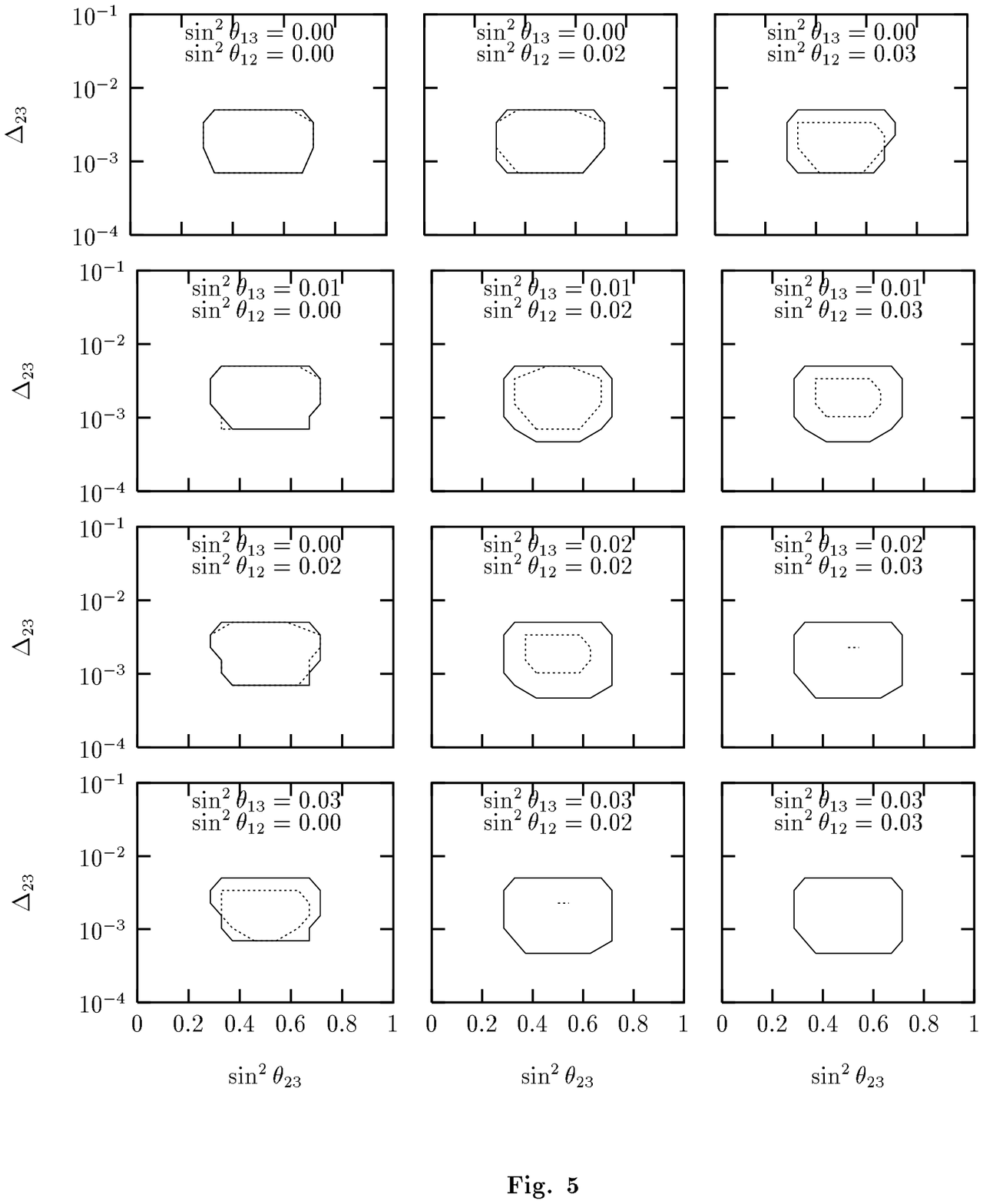}
\end{figure}

\begin{figure}[p]
\epsfxsize 16 cm
\epsfysize 17 cm
\epsfbox[25 151 585 704]{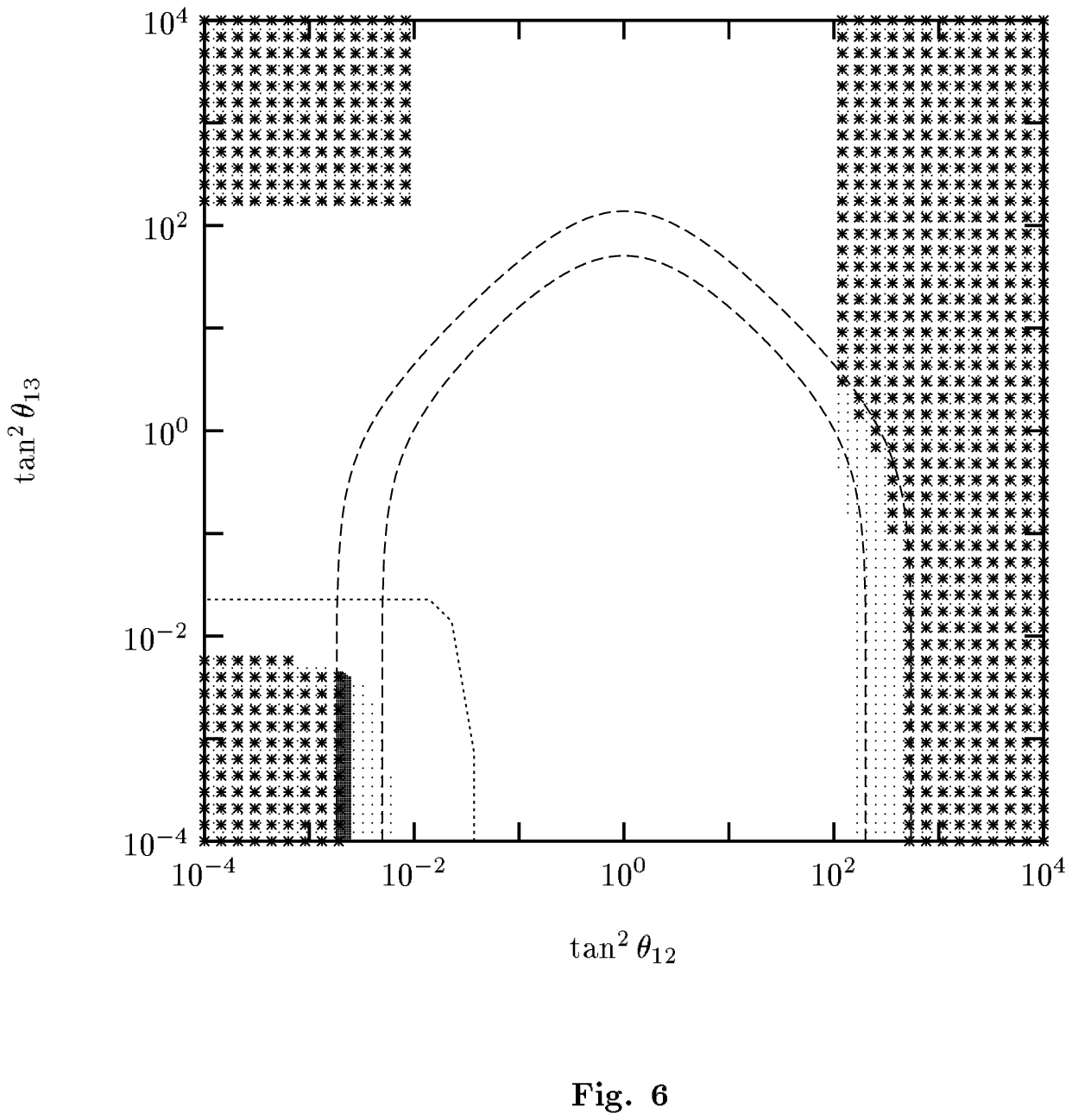}
\end{figure}

\begin{figure}[p]
\epsfxsize 16 cm
\epsfysize 17 cm
\epsfbox[25 151 585 704]{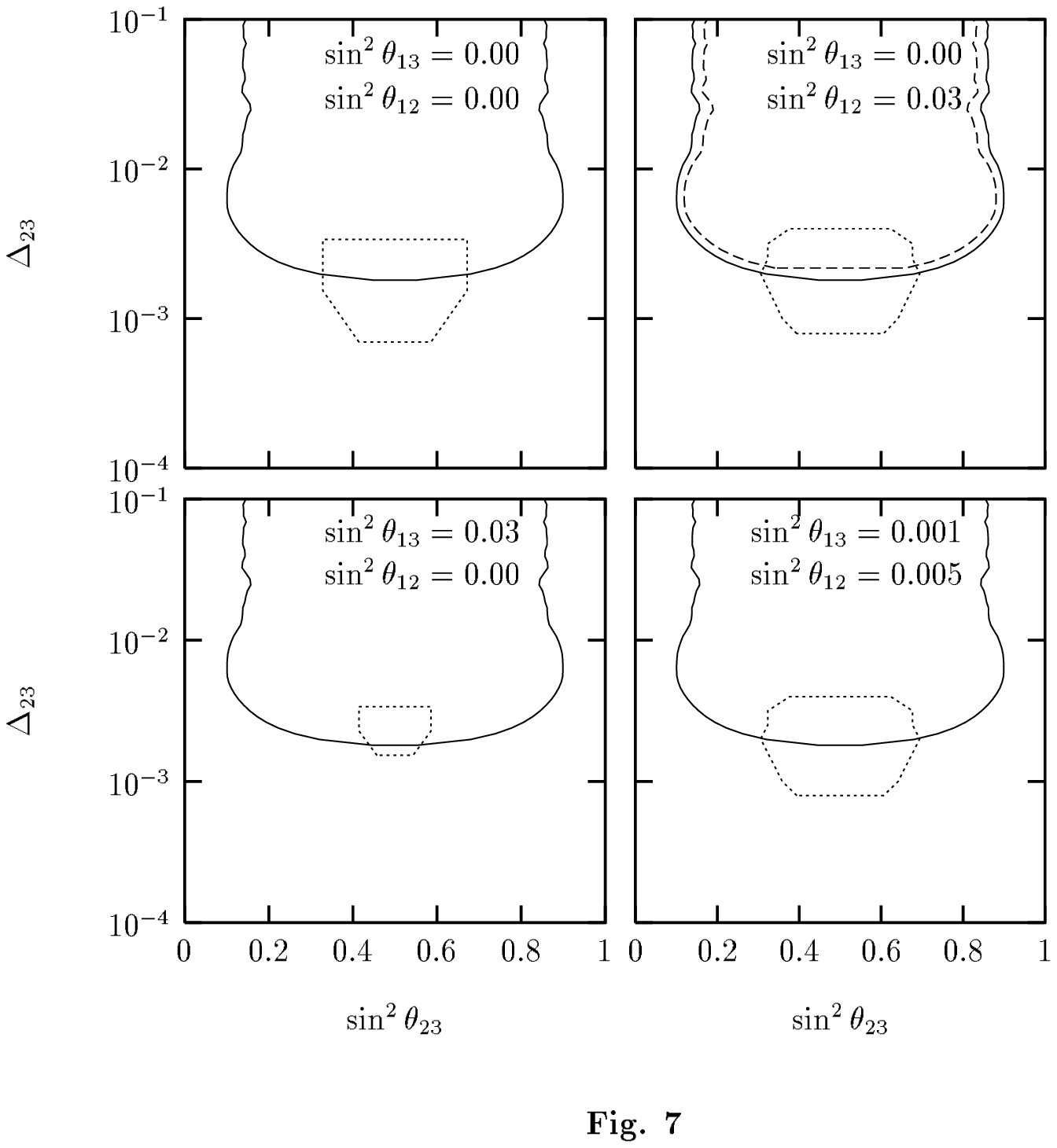}
\end{figure}

\begin{figure}[p]
\epsfxsize 12 cm
\epsfysize 10 cm
\vskip -6.5cm
\hskip 2cm
\epsfbox[25 151 585 704]{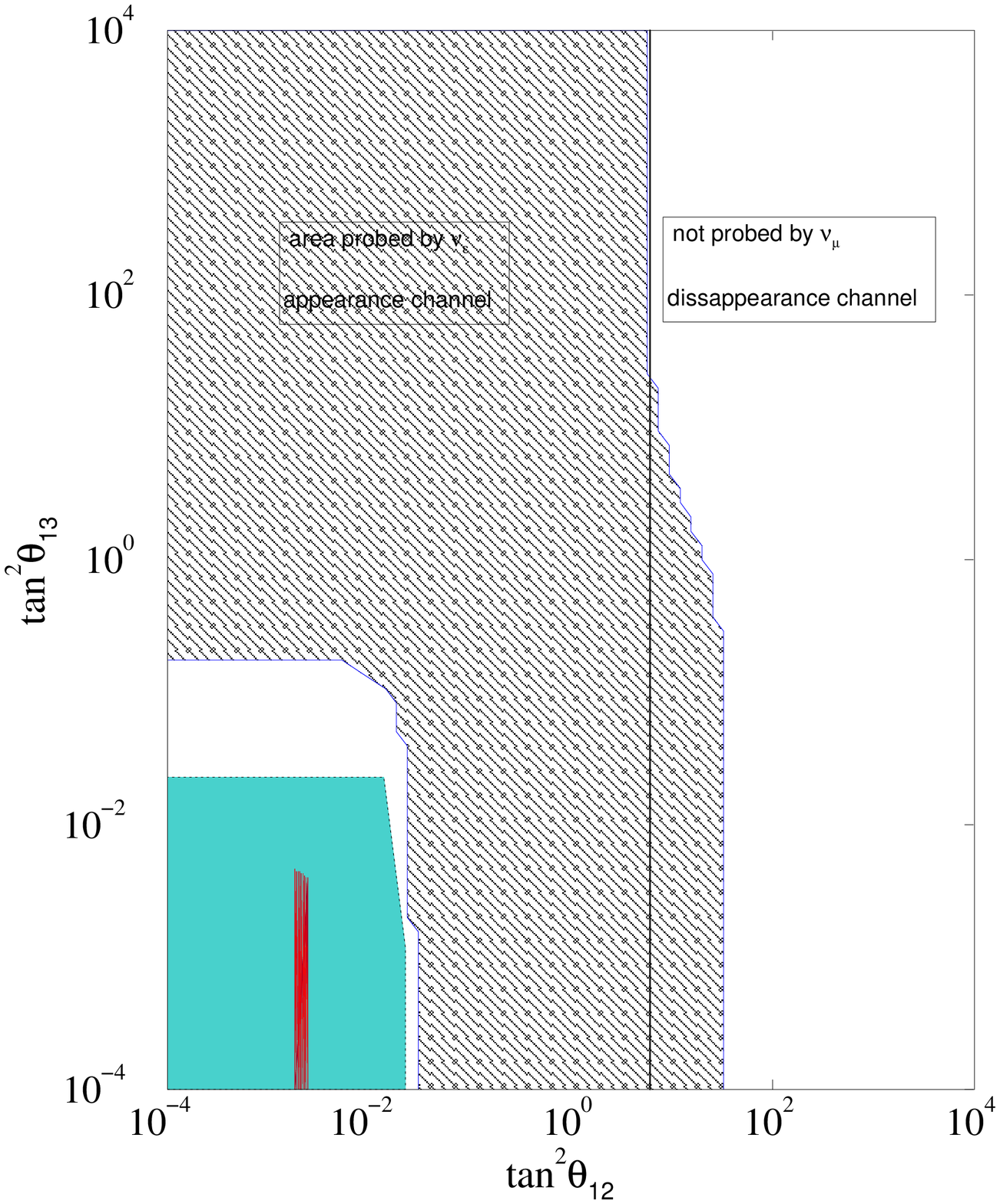}
\begin{center}
\vskip 1.2cm
\bf {Fig. 8}
\end{center}
\end{figure}

\end{document}